\documentclass[%
reprint,
amsmath,amssymb,
aps,
]{revtex4-1}
\usepackage{nicefrac}
\usepackage{multirow}
\usepackage[usenames,dvipsnames]{xcolor}
\usepackage{graphicx}
\usepackage{epsfig}
\usepackage{dcolumn}
\usepackage{bm}
\usepackage[colorlinks=true,citecolor=blue,]{hyperref}
\usepackage{multirow}
\usepackage{array}
\DeclareGraphicsExtensions{.eps,.pdf}
\usepackage{epstopdf}
\newcommand{\ig}[1]{}

\usepackage{array}
\newcolumntype{L}[1]{>{\raggedright\let\newline\\\arraybackslash\hspace{0pt}}m{#1}}
\newcolumntype{C}[1]{>{\centering\let\newline\\\arraybackslash\hspace{0pt}}m{#1}}
\newcolumntype{R}[1]{>{\raggedleft\let\newline\\\arraybackslash\hspace{0pt}}m{#1}}

\begin{document}

\title{Magnetism in tetragonal manganese-rich Heusler compounds}
\author{Lukas~Wollmann,$^{1}$ Stanislav~Chadov,$^{1}$ J\"urgen K\"ubler,$^{2}$ Claudia Felser$^{1}$}
\affiliation{$^1$Max-Planck-Institut f\"ur Chemische Physik fester
        Stoffe,  N\"othnitzer Strasse~40, 01187 Dresden, Germany}
\affiliation{$^2$Institut f\"{u}r Festk\"{o}rperphysik, Technische
        Universit\"{a}t Darmstadt, 64289 Darmstadt, Germany}

\begin{abstract}
A  comprehensive study  of the  total energy  of  manganese-rich Heusler
compounds using density functional  theory is presented. Starting from a
large set of cubic parent systems, the response to tetragonal distortions
is studied in  detail. We single out the systems  that remain cubic from
those  that  most likely  become  tetragonal.  The  driving force  of  the
tetragonal  distortion  and  its  effect  on  the  magnetic  properties,
especially where they deviate from the Slater--Pauling rule, as well as the
trends  in the  Curie temperatures,  are  highlighted. By  means of  partial densities  of states, the  electronic structural changes  reveal the
microscopic origin of the observed trends. We focus our attention on the
magnetocrystalline  anisotropy and  find astonishingly  high  values for
tetragonal   Heusler  compounds   containing  heavy   transition  metals
accompanied  by  low magnetic  moments,  which indicates that these materials are  promising
candidates for  spin-transfer  torque magnetization-switching
applications.

\end{abstract}

\maketitle

\section{Introduction}
\label{sec:Intro}
\begin{figure*}
        \includegraphics[width=1.0\linewidth]{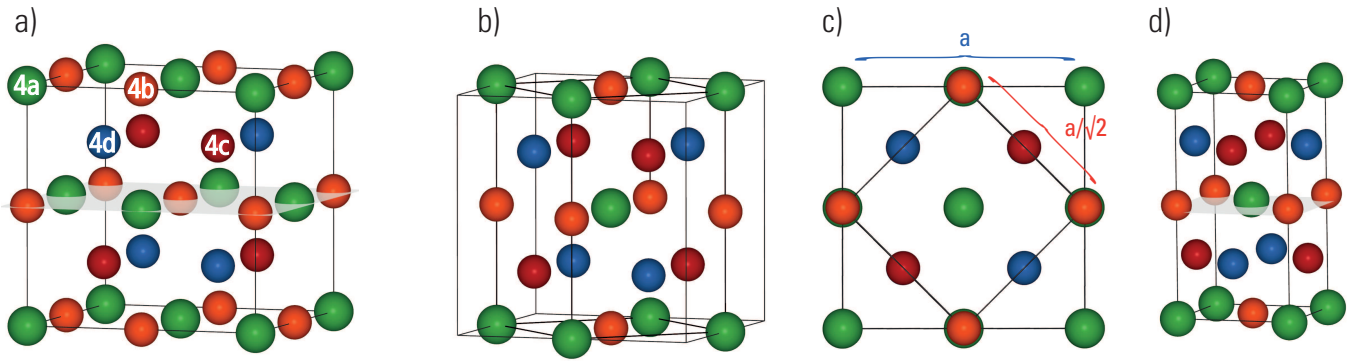}
        \caption{\label{fig0}(Color online) The conversion from a (a) cubic
          Heusler structure to a (b) tetragonal derivative phase is
          displayed in terms of a non-displasive 
          transformation for a system with the general composition of
          $XX'YZ$ ($X$, $X'$, $Y$ - transition metals and $Z$ -
          main-group element, marked as red, orange, blue, and green,
          respectively) within the fcc lattice. (c) Relationship of the
          lattice parameters as ${a_{\rm tet}=a_{\rm cub}/\sqrt{2}}$.}
\end{figure*}
The spintronics community demands materials with uniaxial anisotropy for
spin  transfer torque--random access  memory (STT-RAM)  applications, as
well as  for fundamental  skyrmion-related research or  in the  field of
magnetic shape-memory  alloys~\cite{WCG+2012}. Such materials are needed
to  improve the  functionality and  applicability of  modern  devices or
device  concepts within  the  scope  of  mass production  or  proof  of
practicality.  Addressing this request withing the class of compounds with
Heusler and Heusler-like structures, the task is approached by means of
relatives of familiar systems~\cite{WFC+2015,CDW+2015}. These relatives are the family
of Mn$_2$-based  Heusler compounds,  the famous pioneering  material and
ancestor of which is Mn$_3$Ga~\cite{BFK+2007,WBF+2008}. The uniaxial
magnetocrystalline  anisotropy of Mn$_3$Ga  has  been calculated~\cite{ZYW+2013} and measured~\cite{KRV+2011} several times on different occasions. It is
thought  that  anisotropic materials  such as  these  could constitute  the
foundation for  magnetic racetrack memory as proposed by \textsl{Parkin et al.}~\cite{PHT2008}. The  indispensable
perpendicular magnetic  anisotropy (PMA) in (ultra-)thin structures is
best  controlled by  intrinsic properties  rather than  by shape-  or strain-induced  anisotropy.   As a result, the   perpendicular  orientation  of
magnetization is a  desired  property of  the  material.  Recently,
Mn$_2$-based Heusler systems,  Mn$_2YZ$, were reconsidered as promising
materials.  Thus,  considerable  research   has  been  done  on  related
systems.  Mn$_2$NiGa in particular is  a well-studied  material as  it is
directly  related to  Ni$_2$MnGa, which  has been  the most  studied and  best
understood  ferromagnetic  shape-memory  alloy  since its  discovery~\cite{CCK+1995,UHK+1996}.    
Mn$_2$NiGa,   however,   is    a   ferrimagnetic
shape-memory alloy that is  theoretically linked to Ni$_2$MnGa through a
substitution  series, with  a  transition from  ferro- to  ferrimagnetic
ordering due  to the  increasing manganese content.  In addition  to Mn$_2$NiGa,  other  Mn$_2$-based   Heusler  alloys  have  been  synthesized and
characterized  or  have  been  theoretically  treated,  such as
Mn$_2$CoGa~\cite{AWF+2011,MSK+2011,OKF+2012,KOM+2013,WCK+2014},
Mn$_2$FeGa~\cite{GNW+2013}, and Mn$_3$Ga~\cite{BFK+2007,KRV+2011}. 
In addition to materials in which $Y$ is an atom from
period {\uppercase\expandafter{\romannumeral4\relax}} (or the 3$d$ series),
equivalent systems  with heavier  constituents as $Y$ species  have been
investigated. Among these  were Mn$_2$RuGa~\cite{HAM+2002} (which has
been   found  to have   more   or  less   random   occupation  of   sites),
Mn$_2$RhGa~\cite{KKH+2014} (cubic, disordered), 
Mn$_2$PtGa~\cite{NNC+2013,NNC+2015},   and   Mn$_2$PtIn~\cite{NSW+2013},  in   the
context of large exchange bias effects.\\ 
Although a detailed study  of a single material is an important
task  that  results in  valuable  knowledge,  the  inclusion of  cluster
knowledge    into    a    general    concept    creates    comprehensive
insight. The same intention that  guided our previous work ~\cite{WCK+2014}
motivated  us  to  undertake a  similar  approach in the current study, i.e.,  comprehensively
studying a selected set of systems. We intend to understand the general
trends governing the formation and magnetism of tetragonal materials for
the aforementioned applications.\\  In this paper, we show how
the magnetism and the atomic structure   change withinthe
Mn$_2Y^{\left(3d\right)}$Ga,   Mn$_2Y^{\left(4d\right)}$Ga, and
Mn$_2Y^{\left(5d\right)}$Ga series.  The trigger quantities  causing the
tetragonal distortion, as well as  the consequences of this distortion,
i.e., the magnetocrystalline anisotropy (MCA)~\cite{CDW+2015}, will be
highlighted and placed into an appropriate context. 

\section{Crystal structure}
\label{sec:Cryst} 
Heusler alloys are nowadays  informally divided into two structure
types, the so-called {\it regular} and {\it inverse} types, referring to
the   original   Heusler    compound   Cu$_2$MnAl   as   the   reference
system~\cite{Heu03}. The materials first associated  with the  name of
Fritz Heusler were cubic phases analogous to Cu$_2$MnAl, 
with the stoichiometry $X_2YZ$. From  then on,  similar materials  were
thus labeled ``Heusler compound'', extending the  original definition of Heusler compounds to the family of  Heusler materials  that  incorporates a  variety  of similar
structures. These structures are derived from the original compound, with occupation of  the Wyckoff positions, $8c$, $4b$, and $4a$,  in the spacegroup
(SG) 225 through the introduction  of vacancies or  slight structural
changes. These modifications alter the structure by breaking
the inversion symmetry  when going from a {\it  regular} Heusler phase
to an {\it inverse} or {\it half}-Heusler structure.

\begin{table}[htbp]
\begin{tabular}{L{1.5cm}|C{1cm} C{1cm} |C{1cm}C{1cm}|C{1cm}C{1cm}}
Heusler   &  & SG    &       4$d$    &       4$c$    &       4$b$    &       4$a$\\\hline
regular & L2$_{1}$ & 225 &        Mn      &       Mn      &       $Y$       &       Ga\\
inverse & X$_{a}$  & 216  &  Mn      &       $Y$       &       Mn      &       Ga\\
half    & C1$_{b}$  & 216  &  Mn      &      Mn       &       $\square$      &       Ga\\
\end{tabular}
\caption{\label{tab:StrucRel}The structural relationship of the {\it regular} and {\it inverse} structure types for Mn$_2Y$Ga is shown, where $X$=$X'$=Mn and $Z$=Ga, as compared to Fig.~\ref{fig0}. Only one possible configuration of a {\it half}-Heusler type is listed exemplary.}
\end{table}
Heusler materials  are generally understood  to be intermetallic compounds,
distinguishing  them from  general  intermetallics which  are forming  a
broad  range  of solid  solutions,  with  no  preferred but  statistical
occupation of crystallographic sites. They  are also set apart from other
ionic  or  covalent  compounds  because  Heusler  systems  allow  the
formation of substitutional series  of single sites. However, they maintain the
character of an  ordered compound, and thus they are on  the borderline between
alloys and compounds. Some distinct systems exhibit a tendency to form
alloys, nevertheless. The  general  composition is  given by  $XX'YZ$,
where  the   classical  definition  has  been   widened  to  incorporate
quaternary  materials within  the family  of Heusler  compounds.  In the
representation of SG 216 ($F\bar{4}3m$), the structure contains four highly
symmetric Wyckoff positions: 
$4d$~($\frac{3}{4},\frac{3}{4},\frac{3}{4}$), 
$4c$~($\frac{1}{4},\frac{1}{4},\frac{1}{4}$), 
$4b$~($\frac{1}{2},\frac{1}{2},\frac{1}{2}$), and 
$4a$~($0,0,0$). Depending on
the   occupation  of  the   crystallographic  positions,   two  ordering
possibilities    for    ternary    alloys    (${X=X'}$)    are    obtained
(Table~\ref{tab:StrucRel}).  In   this  study,  gallium   was  chosen  as
the $Z$ element, whereas  one manganese atom,  $X$, occupies position  $4d$. The
second  manganese atom,  $X'$,  and  the other  transition  metal, $Y$,  are
located at $4c$ or $4b$,  respectively, depending on the formation of the
{\it  regular}-   or  {\it  inverse}-type   Heusler material, as  shown  in
Table~\ref{tab:StrucRel}. Other Heusler-related structures~\cite{Suits1976} are  the tetragonal derivatives  of the cubic
parent phases, which  have been widely treated in  the context of magnetic
shape-memory  alloys. The relationship and the  unit cell transformation
between cubic  and tetragonal phases is depicted  in Fig.~\ref{fig0}. It
is seen that a conventional cubic unit cell can be described in terms of
a tetragonal  lattice  exhibiting  a  ${c/a}$ ratio of  $\sqrt{2}$.  
The  cell parameters  are interrelated according  to ${c_{\rm  tet}=c_{\rm cub}}$
,~${a_{\rm tet}=a_{\rm  cub}/\sqrt{2}}$. In  this study, a set of  Mn$_2$-based
materials,  including transition metals  of periods  IV, V,  and VI,     are    considered, namely,    Mn$_2Y^{(3d)}$Ga,    Mn$_2Y^{(4d)}$Ga, and
Mn$_2Y^{(5d)}$Ga.
\begin{figure*}[htbp]
        \centering
        \includegraphics[width=\linewidth]{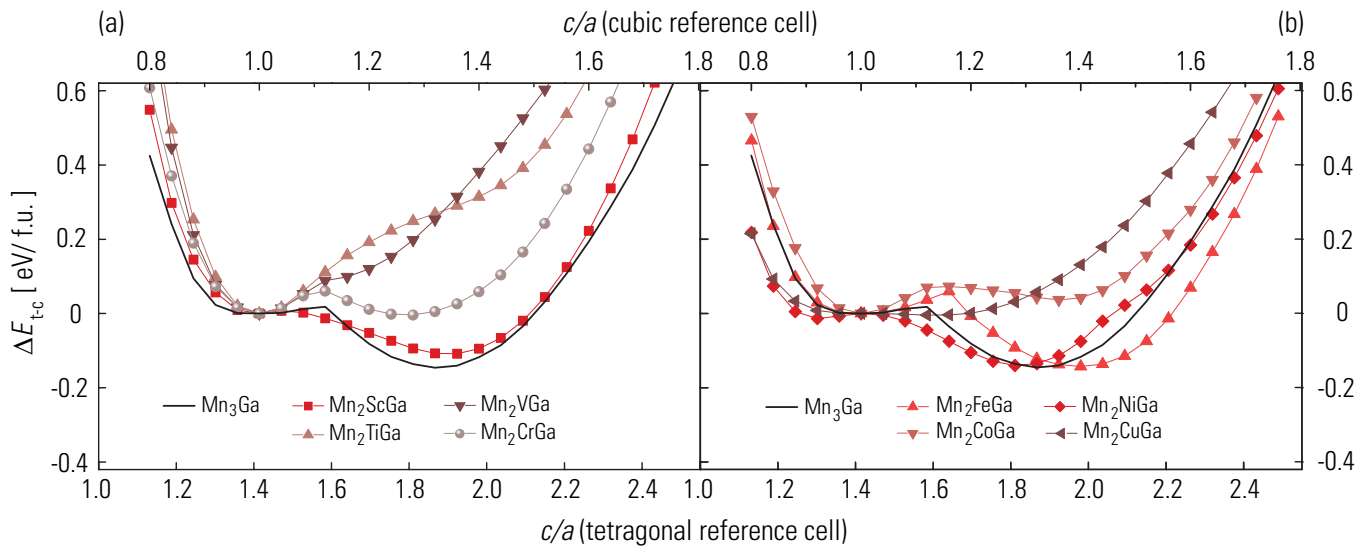}
        \caption{\label{figcoa3d}(Color online) Energetic response to volume-conserving elongations and compressions of the crystal structure along the $c$-axis for the Mn$_2Y^{(3d)}$Ga-series. Panel~(a)~includes systems with a valence electron count ${N_{\rm V}\left(Y\right)\leq7}$ (early transition metals, ETM) and panel (b)~shows the remaining combinations of Mn$_2Y^{(3d)}$Ga with ${N_{\rm V}\left(Y\right)\geq7}$, with so-called late transition metals (LTM).}
\end{figure*}

\section{Computational Details}
\label{sec:CompDet}
The numerical work was done within density functional theory as
implemented in the all-electron FP-LAPW code, \textsc{WIEN2k}
\cite{wien2k}, employing the  generalized gradient approximation (GGA)
in the parametrization of Perdew, Burke, and Enzerhof as
exchange-correlation functional \cite{PBE1996}. The angular momentum
truncation was set to ${l_{\rm max}=9}$ and the number of plane-waves
determined by ${RK_{\rm max}=9}$ to ensure well-converged
calculations. The energy convergence criterion for the self-consistent
field calculations was set to $10^{-5}$~Ry, whereas the charge
convergence was set to $10^{-3}$. All calculations where done on a
${20\times20\times20}$ $k$-mesh. For a set of given $c/a$ ratios, the
volumes were optimized and fitted to the Birch--Murnaghan equation of
state \cite{Birch1947,Murnaghan1944}. From these, the optimal ratio and
volume were obtained and the lattice parameters evaluated. On the basis
of the crystallographic details, complex magnetic properties such as the
exchange parameters and the corresponding Curie temperatures ($T_{\rm
  C}$) were computed by means of the Korringa--Kohn--Rostoker (KKR)
Green's function method as implemented by the Munich SPR-KKR
package~\cite{EKM11}. The angular momentum expansion was truncated for
${l_{\rm max}=3}$ which corresponds to the $f$-wave symmetry. The energy
integration was done on a complex energy mesh with 48 points along the
integration path,  using the Lloyd's formula~\cite{LPSP1972} for an
improved estimate of the Fermi energy. The computation of the exchange
parameters is based on the classical Heisenberg model, 
which was evaluated by means of the real-space
approach~\cite{LKA+1987}. This provides site- and distance-dependent
exchange between sites via infinitesimal rotation of the magnetic 
moments at a particular site in real space. To account for distance
dependence, an appropriate truncation of the cluster radius, $r$, around
each atomic site had been chosen. This radius was set to ${3.5~a}$
lattice spacings to capture even small interactions, 
as the largest contributions to the effective exchange constants are
found for radii smaller than ${1.5~a}$ lattice spacings~\cite{MSR2011b,WCK+2014}.
\begin{figure*}[htbp]
        \includegraphics[width=\linewidth]{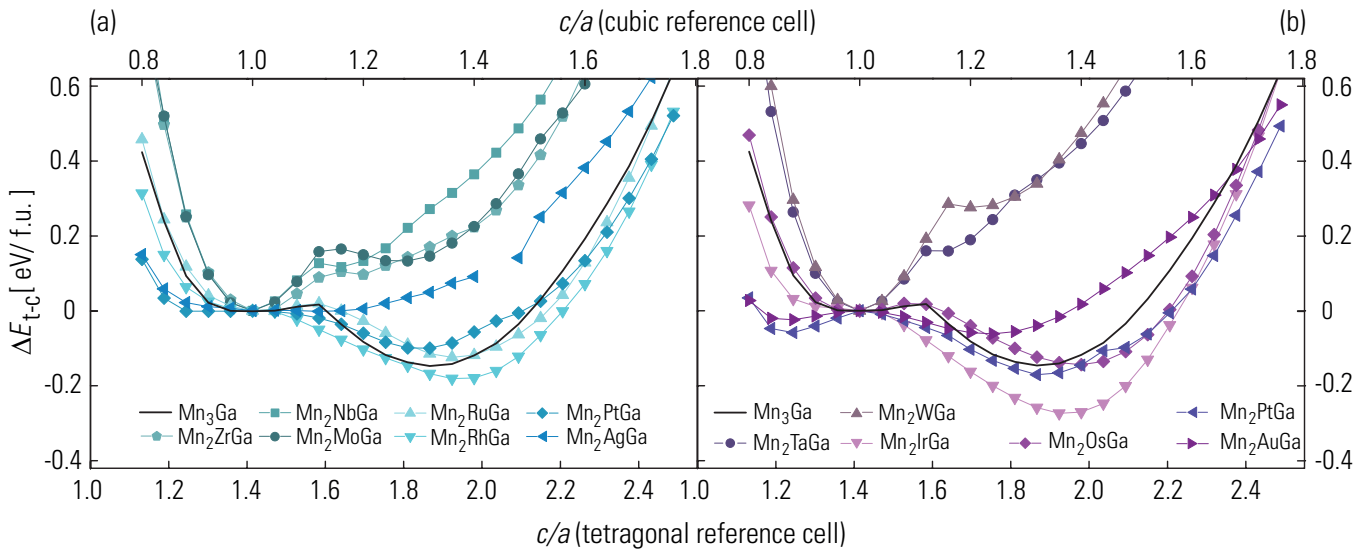}
        \caption{\label{figcoa4d5d}(Color online) Energetic response to volume-conserving elongations and compressions of the crystal structure along the $c$-axis for the (a) Mn$_2Y^{(4d)}$Ga and (b) Mn$_2Y^{(5d)}$Ga series. Systems involving late transition metals and systems exhibiting at least a local tetragonal energy minimum are shown.}
\end{figure*}
\section{Results}
\subsection{Lattice Relaxation}
\label{subsubsec:StrucMagGround}
\begin{figure*}[htbp]
        \includegraphics[width=.75\linewidth]{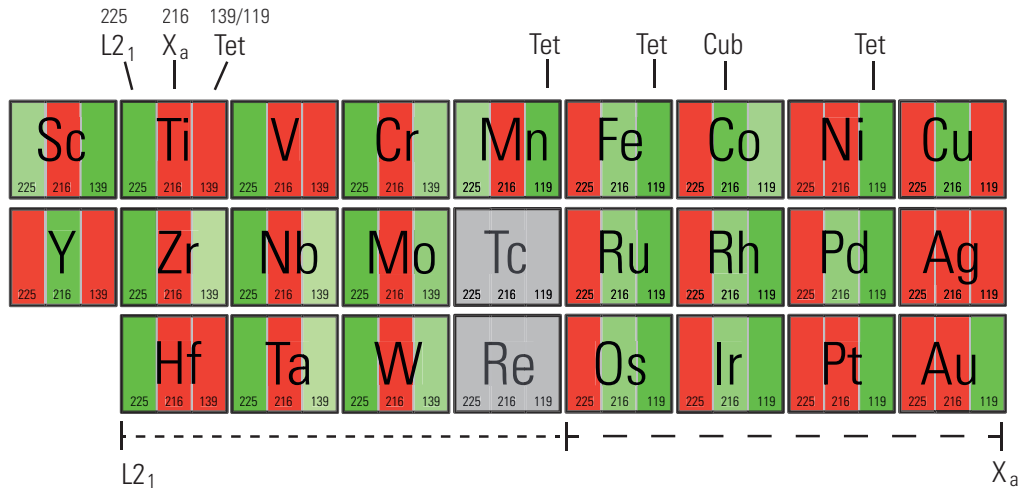}
        \caption{\label{fig:pse}(Color online) Schematic overview of the preferred site occupancy and crystal structure of Mn$_2Y$Ga Heusler compounds. Stable, metastable, and instable lattices are marked by dark-green, light-green, and red subcells, respectively.}
\end{figure*}

\begin{figure}[htbp]
        \centering
        \includegraphics[width=\linewidth]{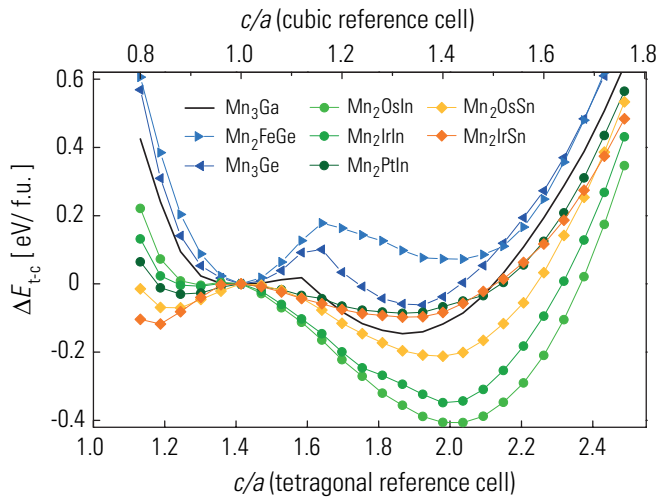} 
        \caption{\label{figcoafurther}(Color online) The energy landscapes, $E(c/a)$, of Mn$_3$Ge, Mn$_2$FeGe, Mn$_2$(Ir,Pt)Sn, and Mn$_2$(Ir,Pt)In are shown.}
\end{figure}

The total energy ${E\left(c/a\right)}$ as a function of the ${c/a}$
ratio was calculated and the results are shown in
Figs.~\ref{figcoa3d}--\ref{figcoafurther}. The energy zero are defined
with respect to the cubic parent compound, and consequently the energy
difference for all phases can be compared easily. 
The case of Mn$_3$Ga is used as a benchmark and is repeatedly plotted in Figs.~\ref{figcoa3d} and \ref{figcoa4d5d}. Table~\ref{tab:LatDat} contains the numerically optimized lattice parameters. The study reveals that a large number of the herein-treated materials are most stable in their respective tetragonal structures, with ${c/a>\sqrt{2}}$.\\
Tetragonally compressed structures are described by $c/a<\sqrt{2}$, whereas tetragonally elongated lattices are characterized by ${c/a>\sqrt{2}}$, as compared to the cubic parent or austenite phase. Elongation occurs with an increase in the length of the $c$-axis, whereas the ${ab}$-plane is compressed, leaving the volume approximately unchanged. In this study, the volume of the unit cell, $V_{\rm cell}$, was optimized in addition to the $c/a$-ratio, and it was found that no significant change occurred for most cases (Table~\ref{tab:Vol}).\\
\begin{table}[htbp]
\begin{ruledtabular}
                \begin{tabular}{lD{.}{.}{4}|lD{.}{.}{4}}
Material & \multicolumn{1}{c}{$C_{\rm t/c}$}  & Material & \multicolumn{1}{c}{$C_{\rm t/c}$}\\\hline
Mn$_2$ScGa      &       5.08      & Mn$_2$OsGa  &       2.32     \\
Mn$_2$CrGa      &       6.11      & Mn$_2$IrGa  &       2.44     \\
Mn$_3$Ga                &       9.08      & Mn$_2$PtGa  &       -3.13    \\
Mn$_2$FeGa      &       1.93      & Mn$_2$AuGa  &       -1.49    \\
Mn$_2$CoGa      &       1.82      & Mn$_2$OsSn  &                    \\
Mn$_2$NiGa      &       -1.09     & Mn$_2$IrSn  &       -1.94    \\
Mn$_2$CuGa      &       2.37      & Mn$_2$PtSn  &       -0.35    \\
Mn$_2$MoGa      &       3.46      & Mn$_2$OsIn  &       -0.36    \\
Mn$_2$RuGa      &       1.63      & Mn$_2$IrIn  &       -1.01    \\
Mn$_2$RhGa      &       1.48      & Mn$_2$PtIn  &       -0.86    \\
Mn$_2$PdGa      &       -2.78     & Mn$_3$Ge            &       3.98     \\
&& Mn$_2$FeGe   &       3.80   \\
                \end{tabular}
\end{ruledtabular}
\caption{\label{tab:Vol}Relative volume change between the austenite and martensite phase in percent, $C_{\rm t/c}=\left(V_{\rm tet}-V_{\rm cub}\right)\cdot 100 / V_{\rm cub}$}
\end{table}
Figs.~\ref{figcoa3d}a and \ref{figcoa3d}b show the $E\left(c/a\right)$ curves of the Mn$_2Y^{(3d)}$Ga series  for early transition metals (ETM, $Y=\rm Sc,~Ti,~V,~Cr$) and late transition metals (LTM, $Y=\rm Mn,~Fe,~Co,~Ni,~Cu$), respectively. A preference for tetragonal structures is seen for materials including LTM for Mn$_2Y^{(3d)}$Ga, with $Y$ being Mn, Fe, or Ni. In the group of ETM systems, Mn$_2$ScGa was found to be tetragonal, which is an exception. The situation for Mn$_2Y^{(4d)}$Ga and Mn$_2Y^{(5d)}$Ga is seen to be similar.\\
In contrast to Mn$_2$CoGa, for which a cubic structure ($c/a=\sqrt{2}$) is preferred, the systems with the same valence electron count ($N_{\rm V}=26$) involving heavier species (ruthenium and osmium) exhibit a global energy minimum for the tetragonal structure.\\
Discontinuities in the $E\left(c/a\right)$ curves are observed for the cases of Mn$_3$Ga and Mn$_2$FeGa, resembling a first-order transition, whereas $E\left(c/a\right)$ for Mn$_2$NiGa is continuous (Fig.~\ref{figcoa3d}b). Strongly composition-dependent modulated martensitic phases \cite{FKC+1996,CSC+1998} and premartensitic phases \cite{ZSW+1995,SVK+1997} have been experimentally produced; in combination with first-principles calculations \cite{ZEE+2003}, the onset of the martensitic transition has been thought to be initiated by a displacement of atomic planes orthogonal to the crystallographic $c$-axis of the tetragonal cell. Experimentally \cite{SVK+1997} and theoretically \cite{ZEE+2003}, it has been shown that phonon softening along $[\zeta,\zeta,0]$ exists in shape-memory materials, and thus the transition has been related to the occurrence of the tetragonal distortion.\\
The approach undertaken in the current study resulted in elongated structures in those cases, which were unstable towards a tetragonal distortion. Additionally, global energy minima for distorted phases were found mostly for the {\it inverse} structures, as depicted in Fig.~\ref{fig:pse}. In contrast, compressed variants possessing a global energy minimum were not observed in the Mn--Ga Heusler family. 
The calculated lattice data and the figures shown (Figs.~\ref{figcoa3d}--\ref{figcoa4d5d}) lead to the conclusion that stable tetragonal structures may only be formed in the series of Mn$_2Y$Ga that includes LTMs, making them derivatives of \textit{inverse} Heusler systems. The information obtained from the lattice optimization in terms of the relative positions of the energy minima is condensed in Fig.~\ref{fig:pse}, which gives the preferred crystal structure  visualized  in the manner of the periodic table.\\
Each compound in the family of Mn$_2Y$Ga materials shown in this figure is depicted by one cell that symbolizes a transition metal $Y$ of the $3d,~4d$, or $5d$ series. The corresponding cell is built up by three subcells, which represent the two variants of chemical coordination (the first or second subcell) and the existence of a global energy minimum for $c/a \neq \sqrt{2}$ (third subcell). The color code symbolizes the energy level of a configuration on the energy landscape relative to one another: dark green - global minimum, light green - local minimum, red - no minimum. In cases in which the investigated materials did not exhibit a cubic minimum, the first two subcells are understood to be the type of coordination around the 4$d$ crystallographic site: the symmetry of the coordinating shell is either centrosymmetric (in relation to SG 225) or non-centrosymmetric (in relation to SG 216). As shown in a preceding publication  on the cubic variants \cite{WCK+2014}, systems involving ETMs adopt the L2$_1$-type structure, whereas compounds containing LTMs are found to have the {\it inverse} Heusler structure (X$_a$-type).\\
Inspection of Fig.~\ref{fig:pse} in combination with Figs.~\ref{figcoa3d} and \ref{figcoa4d5d} reveals interesting details, such as the fact that tetragonal derivative phases of cubic Heusler alloys, implying that a global energy minimum is present, are observed only for a valence electron count of $N_{V}\geq 24$. It is also clearly seen that the onset of the formation of tetragonally elongated structures evolves over the periods from left to right and from lower to higher $N_V$.\\
This leads  to the question of the mechanism behind this distortion, which we are going to approach in Sec.~\ref{subsubsec:AnaElStruct}.\\
The lattice parameters for all cubic compounds are found within a range of $\Delta a_{c/\sqrt{2}} =0.35\ \rm\AA$ for the Ga series. The values increase from the borders of the series towards the middle of the range. The same behavior is found for tetragonal compounds, where the range spans from $a_{\rm tet} = 3.68$ to $4.11$ ($\Delta a_{\rm tet}=0.43\ \rm\AA$) and ${c_{\rm tet}=6.91}$ to $7.48$ (${\Delta c_{\rm tet}=0.54\ \rm\AA}$), whereas Mn$_2$ScGa is an exception. The $c/a$ coordinate exhibits inverted behavior, decreasing from the middle to the left and right borders of the series. The overall similarity of the lattice data opens the possibility of intermixability with each other  and thus tunability of the whole class of materials. Therefore, the magnetization may be adjusted over a large range, allowing for the formation of tetragonal compensated ferrimagnets via adequate substitution.\\
Having discussed the structural trends as a function of the valence electron change in the $d$-electron system, the effects of the  variation  of the main group element, $Z$, from Ga to Ge, In, and Sn in a small sample of compounds [Mn$_2$(Mn,Fe)Ge, Mn$_2$(Os,Ir,Pt)In, and Mn$_2$(Os,Ir,Pt)Sn] are briefly discussed. The corresponding data, including those for Mn$_3$Ga, are graphed in Fig.~\ref{figcoafurther}.\\
Fig.~\ref{figcoafurther} reveals that the placement of In and Sn at the $Z$ position leads to the emergence of a deep energy minimum for Mn$_2$IrSn with ${c/a<\sqrt{2}}$ and even deeper minima for elongated phases with large $c/a$-ratios of Mn$_2$OsIn, Mn$_2$IrIn and Mn$_2$OsSn  whose lattice parameters resemble those of layered structures. Heusler alloys are often interpreted in terms of a rigid, band model-like approach, and an electron-filling scheme is employed for the prediction of magnetic moments in Co$_2$-based alloys. Interpretation of the tetragonal instabilities using such an approach leads to the assumption of Mn$_3$Ge ($N_{\rm V}=25$) behaving analogously to Mn$_2$FeGa (${N_{\rm V}=25}$) and of Mn$_2$FeGe (${N_{\rm V}=26}$) being similar to Mn$_2$CoGa (${N_{\rm V}=26}$). A comparison of the corresponding $c/a$ curves demonstrates that this is approximately true and thus that isoelectronicity is an appropriate concept in the chemistry and physics of Heusler materials.\\
Although our calculations agree well with those of previous works \cite{CSR+2013}, comparison with the experimental data of Mn--Ni--Ga systems exhibits an interesting discrepancy. The deviation between experiment and theory was traced back to the deviating structural model in terms of the occupation  of the involved sites. Neutron diffraction studies on Mn$_2$NiGa  highlighted the fact that the order is different from the expected $X_a$-type in the austenite phase. Thus, the chemical formula reads as (Mn,Ni)$_2$MnGa and is called L2$_{1b}$-type because the point symmetry includes inversion symmetry through random occupation of $4d$ and $4c$ sites with Ni and Mn \cite{BKN+2010}. Similarly in Mn$_2$FeGa, a deviation from theory was found in an experimental study because the site occupation was expected to differ from the perfect MnFeMnGa ordering \cite{GNW+2013}. Similar issues are present for the Mn$_2Y^{(4d)}$Ga series. From available data, including neutron diffraction studies, the site occupancies have been clarified. Several authors found that members of the Mn$_2$Ru$Z$ series ($Z=\rm Ga,~Ge,~Si,~Sn$) and Mn$_2$Rh$Z$ series ($Z=\rm ~Ga$)~\cite{HAM+2002,KKH+2014,EKH+2012} not to be tetragonal under the respective reaction conditions. In contrast, they have been realized as cubic Heusler alloys exhibiting a strong degree of anti-site disorder, which has been characterized as an alloying tendency~\cite{KKH+2014}.\\
Orthorhombic deformations of the unit cell have been observed in some systems such as Mn$_2$NiGa \cite{CSR+2013}. Thus, the restriction to tetragonal distortions and ordered compounds in conducted studies leads to a simplified description of these materials. Nevertheless, a general understanding can be obtained in approaching the Mn$_2$-based Heusler systems through this ansatz. In future studies that aim to predict ground-state structures and magnetic configurations, the parameter space for the atomic sites and relaxations paths for the electronic and spin degrees of freedom has to be enlarged and restrictions that are widely used have to be dropped. As a consequence of the applied restrictions, disorder effects were not incorporated into this study, thus leaving open any explanations of the deviations from experimental results.


\begin{table*}[htbp]
        \caption{\label{tab:LatDat}Calculated lattice parameters of the cubic parent phases and corresponding tetragonal structures, in comparison with existing literature data. The lattice parameters are given in {\AA}ngstroms. The saturation magnetization data, $M_S$, are given in $\mu_{\rm B}/~\rm f.u.$ The listed literature data refer to theoretical ($x$t) or experimental ($x$e) investigation of the mentioned cubic (c$y$) or tetragonal (t$y$) phases, respectively.}
        \begin{ruledtabular}
                \begin{tabular}{lc|cR{6mm}C{6mm}C{6mm}D{.}{.}{2}|C{7mm}C{7.5mm}C{7.5mm}C{7.5mm}D{.}{.}{2}|ccccclc}
 	&	$N_{V}$ 	&	SG 	&	 $c/a$   	&	 $a_{c}$ 	&	$a_{t}$ 	&	\multicolumn{1}{c}{$M_{S}$} 	&	SG	&	$c/a$   	&	 $a_{t}$ 	&	 $c_{t}$ 	&	      \multicolumn{1}{c}{$M_{S}$} 	&	SG	&	 $c/a$ 	&	 $a_{t}$	&	 $c_{t}$ 	&	 	&	ordering 	&	Lit.\\\hline	
 	Mn$_2$ScGa      	&	20	&	225	&	       $\sqrt{2}$      	&	6.15	&	4.35	&	-4.00	&	139	&	1.94	&	3.98	&	7.70	&	-5.36	&	               	&	                                       	&	                	&	                       	&	                	&	               	&	\\	
 	Mn$_2$TiGa      	&	21	&	225	&	       $\sqrt{2}$      	&	5.95	&	4.21	&	-2.97	&	139	&	                       	&	                       	&	                       	&	                       	&	225	&	        $\sqrt{2}$     	&	                	&	5.95	&	ct	&	        MnMnTiGa 	&	 [\onlinecite{MSR2011}]\\	
 	Mn$_2$VGa       	&	22	&	225	&	       $\sqrt{2}$      	&	5.82	&	4.12	&	-1.98	&	139	&	                       	&	                       	&	                       	&	                       	&	225	&	        $\sqrt{2}$     	&	                	&	5.91	&	ce	&	        MnMnVGa         	&	       [\onlinecite{KKM+2008}],[\onlinecite{BvE1981}]\\	
 	Mn$_2$CrGa      	&	23	&	225	&	       $\sqrt{2}$      	&	5.76	&	4.07	&	-1.00	&	139	&	1.82	&	3.82	&	6.95	&	-2.75	&	225	&	$\sqrt{2}$ 	&   	&	5.77   	&	ct 	&	 MnMnCrGa            	&[\onlinecite{LZL+2008}]             \\	
 	Mn$_3$Ga        	&	24	&	225	&	       $\sqrt{2}$      	&	5.82	&	4.12	&	0.01	&	139	&	1.82	&	3.90	&	7.08	&	-1.89	&	139	&	1.77	&	3.77	&	7.16	&	tt	&	        MnMnMnGa       	&	        [\onlinecite{BFK+2007}]\\	
 	&	               	&	               	&	               	&	               	&	               	&	               	&	               	&	               	&	               	&	               	&	               	&	139	&	1.82	&	3.90	&	7.09	&	        te      	&	        MnMnMnGa       	&	        [\onlinecite{BFK+2007}]                                                 \\	
 	Mn$_2$FeGa      	&	25	&	216	&	       $\sqrt{2}$      	&	5.79	&	4.09	&	1.03	&	119	&	1.98	&	3.68	&	7.29	&	-0.78	&	119	&	1.90	&	3.79	&	7.19	&	        te      	&	        Mn(Fe,Mn)$_2$Ga        	&	        [\onlinecite{GNW+2013}]\\	
 	Mn$_2$CoGa      	&	26	&	216	&	       $\sqrt{2}$      	&	5.78	&	4.09	&	2.00	&	119	&	1.93	&	3.71	&	7.14	&	0.17	&	216	&	        $\sqrt{2}$     	&	                	&	5.86	&	        ce     	&	        MnCoMnGa        	&	        [\onlinecite{LDL+2008}]\\	
 	Mn$_2$NiGa      	&	27	&	216	&	       $\sqrt{2}$      	&	5.85	&	4.14	&	1.18	&	119	&	1.82	&	3.79	&	6.91	&	1.00	&	216	&	        $\sqrt{2}$     	&	                	&	5.91	&	        ce     	&	                	&	        [\onlinecite{GCL+2005}]\\	
 	&	               	&	               	&	               	&	               	&	               	&	               	&	               	&	               	&	               	&	               	&	               	&	119	&	1.72	&	3.91	&	6.70	&	        te      	&	               	&	        [\onlinecite{GCL+2005}]\\	
 	&	               	&	               	&	               	&	               	&	               	&	               	&	               	&	               	&	               	&	               	&	               	&	225	&	        $\sqrt{2}$     	&	                	&	5.94	&	        ce     	&	        (Mn,Ni)$_2$MnGa         	&	        [\onlinecite{BKN+2010}]\\	
 	&	               	&	               	&	               	&	               	&	               	&	               	&	               	&	               	&	               	&	               	&	               	&	139	&	1.74	&	3.89	&	6.77	&	        te      	&	        (Mn,Ni)$_2$MnGa        	&	        [\onlinecite{BKN+2010}]\\	
 	&	               	&	               	&	               	&	               	&	               	&	               	&	               	&	               	&	               	&	               	&	               	&	216	&	        $\sqrt{2}$     	&	                	&	5.84	&	        ct     	&	        MnNiMnGa        	&	        [\onlinecite{CSR+2013}]\\	
 	Mn$_2$CuGa      	&	28	&	216	&	       $\sqrt{2}$      	&	5.94	&	4.20	&	0.33	&		&		&		&		&		&	216	&	        $\sqrt{2}$     	&	                	&	5.94	&	        ct     	&	        MnCuMnGa        	&	        [\onlinecite{CSR+2013}]\\\hline	
 	Mn$_2$ZrGa      	&	21	&	225	&	       $\sqrt{2}$      	&	6.14	&	4.34	&	-3.00	&	119	&	               	&	               	&	               	&	               	&	               	&	               	&	               	&	               	&	               	&	               	&	                                                       \\	
 	Mn$_2$NbGa      	&	22	&	225	&	       $\sqrt{2}$      	&	6.00	&	4.24	&	-2.00	&	119	&	               	&	               	&	               	&	               	&	               	&	               	&	               	&	               	&	               	&	               	&	                                                       \\	
 	Mn$_2$MoGa      	&	23	&	225	&	       $\sqrt{2}$      	&	5.91	&	4.18	&	-1.01	&	119	&	1.81	&	3.89	&	7.04	&	-2.99	&	               	&	               	&	               	&	               	&	               	&	               	&	                                                       \\	
 	Mn$_2$RuGa      	&	25	&	216	&	       $\sqrt{2}$      	&	5.96	&	4.22	&	1.03	&	119	&	1.96	&	3.80	&	7.45	&	-0.24	&	216	&	        $\sqrt{2}$ 	&	                	&	6.00	&	ce 	&	 (Mn$_{\frac{2}{3}}$,Ru$_{\frac{1}{3}}$)$_3$Ga 	&	[\onlinecite{HAM+2002}]\\	
 	Mn$_2$RhGa      	&	26	&	216	&	       $\sqrt{2}$      	&	5.98	&	4.23	&	1.64	&	119	&	1.94	&	3.82	&	7.43	&	0.10	&	225	&	        $\sqrt{2}$     	&	                	&	6.03	&	ce	&	        (Mn,Rh)$_2$MnGa         	&	        [\onlinecite{KKH+2014}]\\	
 	Mn$_2$PdGa      	&	27	&	216	&	       $\sqrt{2}$      	&	6.12	&	4.33	&	0.55	&	119	&	1.84	&	3.93	&	7.23	&	0.93	&	               	&	               	&	               	&	               	&	               	&	               	&	                                                       \\	
 	Mn$_2$AgGa      	&	28	&	216	&	       $\sqrt{2}$      	&	6.22	&	4.40	&	0.34	&	119	&	               	&	               	&	               	&	               	&	               	&	               	&	               	&	               	&	               	&	               	&	                                                       \\\hline	
 	Mn$_2$HfGa      	&	21	&	225	&	       $\sqrt{2}$      	&	6.12	&	4.33	&	-2.99	&	119	&	               	&		&	               	&	               	&	               	&	               	&	               	&	               	&	               	&	               	&	               	                                                       \\
 	Mn$_2$TaGa      	&	22	&	225	&	       $\sqrt{2}$      	&	6.00	&	4.24	&	-1.99	&	119	&	               	&		&	               	&	               	&	               	&	               	&	               	&	               	&	               	&	               	&	               	                                                       \\
 	Mn$_2$WGa       	&	23	&	225	&	       $\sqrt{2}$      	&	5.92	&	4.19	&	-0.94	&	119	&	               	&	               	&	               	&	               	&	               	&	               	&	               	&	               	&	               	&	               	&	                                                       \\	
 	Mn$_2$OsGa      	&	25	&	216	&	       $\sqrt{2}$      	&	5.95	&	4.21	&	1.02	&	119	&	1.97	&	3.80	&	7.48	&	-0.28	&	               	&	               	&	               	&	               	&	               	&	               	&	                                                       \\	
 	Mn$_2$IrGa      	&	26	&	216	&	       $\sqrt{2}$      	&	5.97	&	4.22	&	2.00	&	119	&	1.95	&	3.83	&	7.44	&	0.11	&	               	&	               	&	               	&	               	&	               	&	               	&	                                                       \\	
 	Mn$_2$PtGa      	&	27	&	216	&	       $\sqrt{2}$      	&	6.13	&	4.33	&	0.44	&	119	&	1.87	&	3.91	&	7.31	&	0.75	&	119	&	1.38	&	4.37	&	6.05	&	        te      	&	        MnPtMnGa       	&	       [\onlinecite{NNC+2013}]\\	
 	Mn$_2$AuGa      	&	28	&	216	&	       $\sqrt{2}$      	&	6.26	&	4.42	&	0.19	&	119	&	1.73	&	4.11	&	7.13	&	0.14	&	               	&	               	&	                	&	               	&	               	&	                	&	                                                         \\[2mm]\hline	
 	Mn$_2$OsSn      	&	25	&	216	&	       $\sqrt{2}$      	&	6.21	&	4.39	&	1.50	&	119	&	1.95	&	3.97	&	7.75	&	-0.02	&	               	&	               	&	                	&	               	&	               	&	                	&	                                                       \\	
 	Mn$_2$IrSn      	&	26	&	216	&	       $\sqrt{2}$      	&	6.31	&	4.46	&	0.41	&	119	&	1.91	&	4.01	&	7.67	&	0.45	&	119	&	1.54	&	4.29	&	6.59	&	        te      	&	        MnIrMnSn 	&	 [\onlinecite{OCN+2014}]\\	
 	Mn$_2$PtSn      	&	27	&	216	&	       $\sqrt{2}$      	&	6.39	&	4.52	&	0.19	&	119	&	1.81	&	4.15	&	7.52	&	-0.02	&	119	&	1.35	&	4.51	&	6.08	&	        te      	&	        MnPtMnSn       	&	        [\onlinecite{WCG+2012}]\\\hline	
 	Mn$_2$OsIn      	&	25	&	216	&	       $\sqrt{2}$      	&	6.26	&	4.43	&	0.62	&	119	&	2.02	&	3.93	&	7.93	&	-0.27	&	               	&	               	&	                	&	               	&	               	&	                	&	                                                       \\	
 	Mn$_2$IrIn      	&	26	&	216	&	       $\sqrt{2}$      	&	6.30	&	4.45	&	0.68	&	119	&	1.98	&	3.97	&	7.85	&	0.07	&	               	&	               	&	                	&	               	&	               	&	                	&	                                                       \\	
 	Mn$_2$PtIn      	&	27	&	216	&	       $\sqrt{2}$      	&	6.37	&	4.51	&	0.31	&	119	&	1.84	&	4.12	&	7.57	&	0.38	&	119	&	1.57	&	4.32	&	6.77	&	        te      	&	        MnPtMnIn       	&	        [\onlinecite{NSW+2013}]\\\hline	
 	Mn$_3$Ge        	&	25	&	216	&	       $\sqrt{2}$      	&	5.76	&	4.07	&	1.01	&	119	&	1.90	&	3.74	&	7.10	&	-0.98	&	225	&	1.91	&	3.81	&	7.26	&	        te      	&	        MnMnMnGe       	&	        [\onlinecite{KBR+2012}]\\	
 	&		&		&		&		&		&		&		&		&		&		&		&	225	&	1.90	&	3.75	&	7.12	&	        tt      	&	        MnMnMnGe       	&	      [\onlinecite{ZYW+2013}]\\	
 	Mn$_2$FeGe      	&	26	&	216	&	       $\sqrt{2}$      	&	5.73	&	4.05	&	2.01	&	119	&	2.05	&	3.63	&	7.42	&	-0.06	&	216	&	        $\sqrt{2}$     	&	                	&	5.80	&	        ct     	&	        MnFeMnFe        	&	        [\onlinecite{LZZ+2008}]\\	
                \end{tabular}
            \end{ruledtabular}
        \end{table*}


\subsubsection{Analysis of the Densities of States}
\label{subsubsec:AnaElStruct}
\begin{figure}[htbp]
\begin{flushleft}
\includegraphics[width=.47\textwidth]{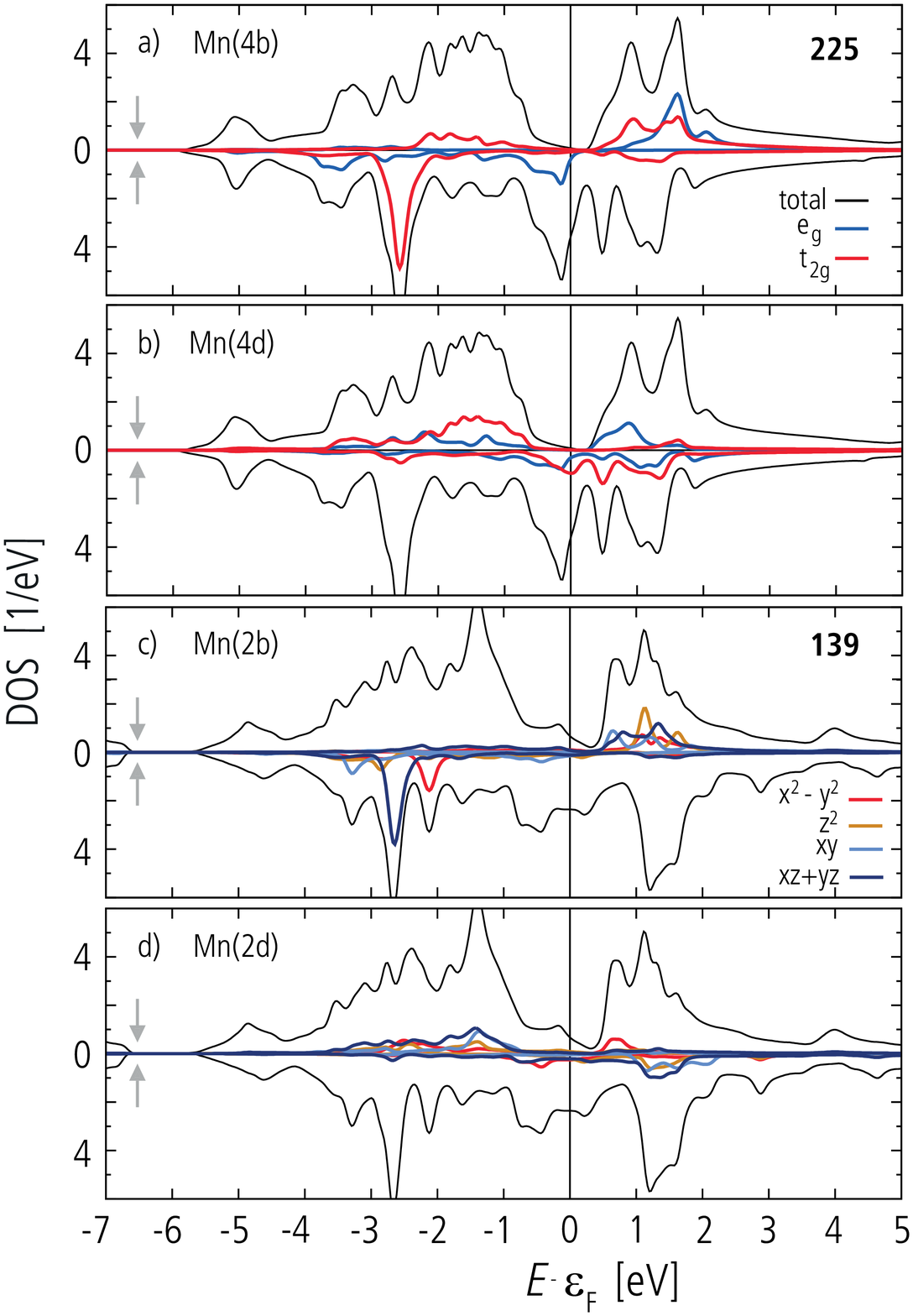}
\end{flushleft}
\caption{\label{fig:DosMn3Ga}(Color online) DOS of cubic (SG 225) and tetragonal (SG 139) Mn$_3$Ga.}
\end{figure}

\begin{figure}[htbp]
\begin{flushleft}
\includegraphics[width=.47\textwidth]{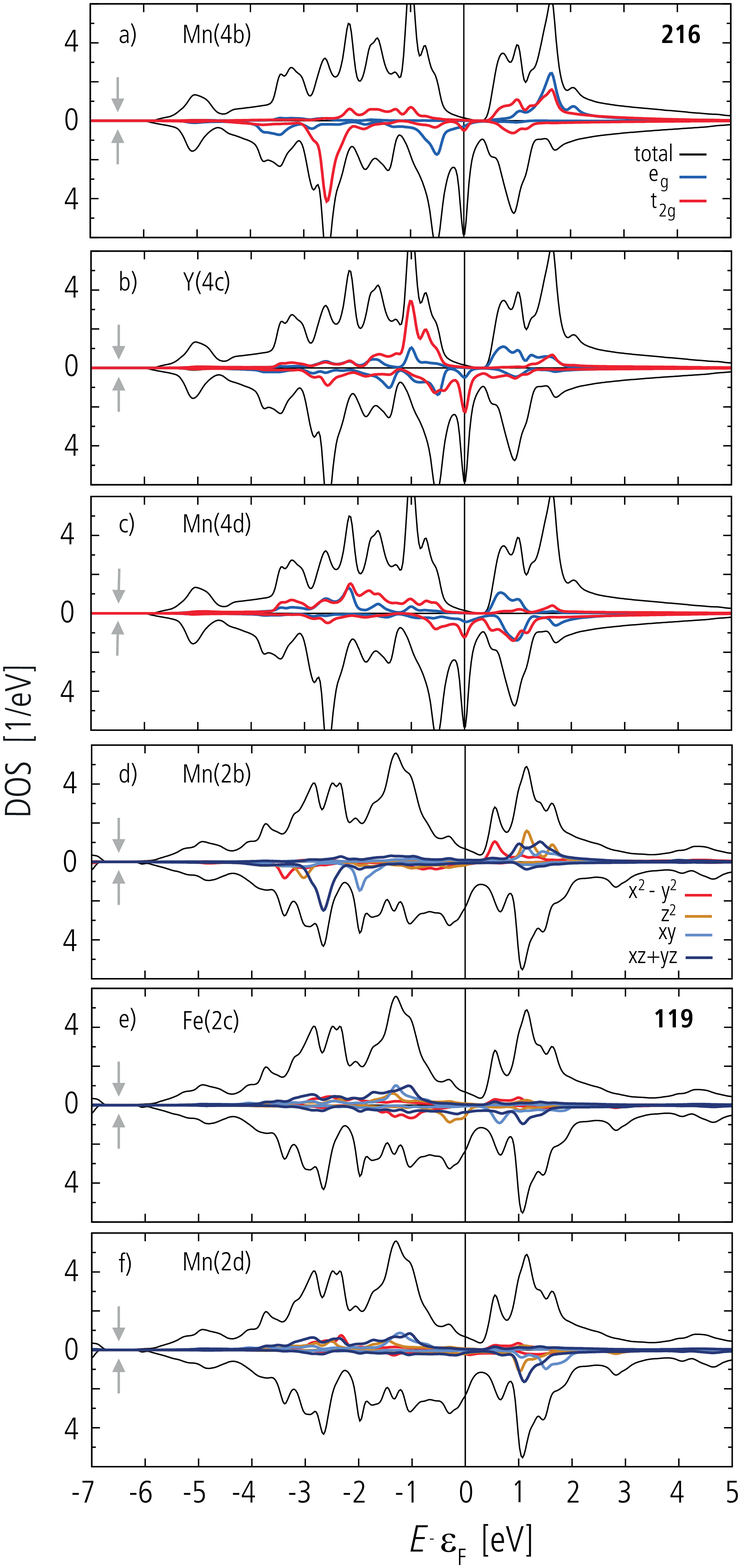}
\end{flushleft}
\caption{\label{fig:DosMn2FeGa}(Color online) DOS of cubic (SG 216) and tetragonal (SG 119) Mn$_2$FeGa.}
\end{figure}

\begin{figure}[htbp]
        \begin{flushleft}
                \includegraphics[width=.47\textwidth]{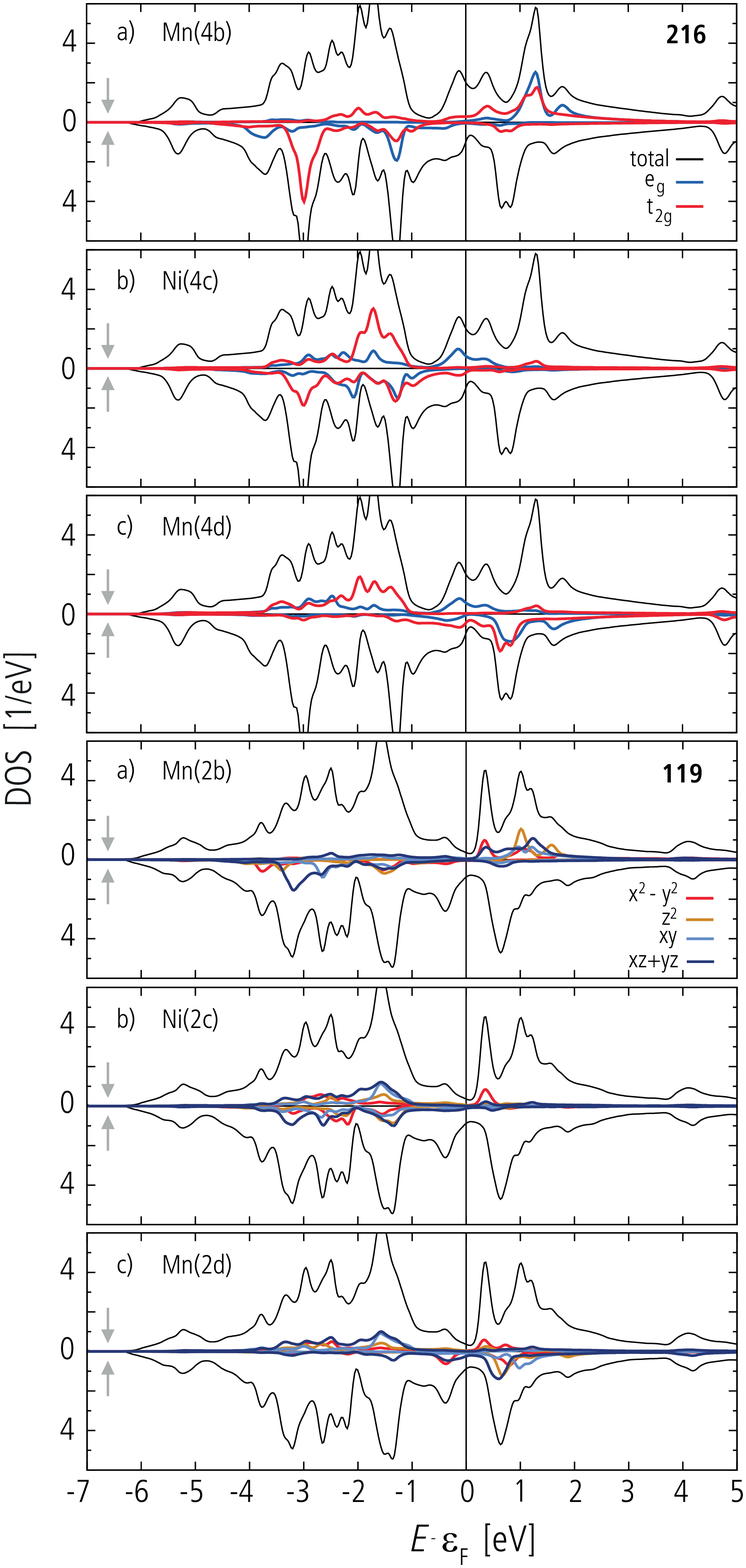}
        \end{flushleft}
        \caption{\label{fig:DosMn2NiGa}(Color online) DOS of cubic (SG 216) and tetragonal (SG 119) Mn$_2$NiGa.}
\end{figure}

Various  attempts to explain the instability of the cubic phase have been given in the literature using models such as the band Jahn--Teller effect \cite{WBF+2008,GNW+2013}, anomalous phonon vibrations \cite{ZER+2005,EBK+2006}, and Fermi-surface nesting \cite{BBS+2007}. These different approaches describe the same behavior, i.e., the instability of the cubic phase, from different perspectives and extracting different types of information. Commonly, the densities of states (DOSs) of related austenite and martensite are compared and contrasted marking the starting and the endpoint of the transition.
As example for  the Mn$_2Y^{(3d)}Z$ series, the DOSs of Mn$_3$Ga, Mn$_2$FeGa, and Mn$_2$NiGa are shown in Figs.~\ref{fig:DosMn3Ga}--\ref{fig:DosMn2NiGa}.
For the cubic variants, the partial DOSs (PDOSs) are shown in their corresponding projections on the sites and in terms of projections on the irreducible representations. The peaks in the PDOS of Mn$(4b)$ are well separated on the energy scale. On the one hand, this separation is due to the strong crystal field splitting of the Mn$(4b)$ $d$-states, where the occupied $e_{g}$ states are located in a range between $-4$ and $-3~\rm eV$, whereas the $t_{2g}$ PDOS is found between $-1~\rm eV$ and the Fermi edge, $\varepsilon_F$. On the other hand, the separation of occupied and empty states follows from the strong exchange split of the Mn$(4b)$ $d$-states, in contrast to the $d$-states of $Y(4c)$ and Mn$(4d)$, which are found to be more widely dispersed even though the majority and minority states are separated owing to exchange splitting.
The PDOSs of Mn$(4d)$ and $Y(4c)$ are strongly dispersed, with
considerable overlap of the spectral weight between the $t_{2g}$ and
$e_{g}$ states in the majority channel, whereas the minority channel is
gapped, with $t_{2g}$ characterizing the lower boundary and $e_{g}$
comprising the upper boundary of the gap. As the DOS is gapped in the
minority spin channel the study of the tetragonal system is strongly facilitated as the states at the Fermi edge in the majority channel mostly constitute the origin of the tetragonal distortion. 
The majority PDOSs in the range of $-5~\rm eV$ up to $\varepsilon_F$ exhibit a characteristically shaped peak structure. From Mn$_3$Ga to Mn$_2$CoGa, the majority spin channel (lower panels of Figs.~\ref{fig:DosMn3Ga}a--\ref{fig:DosMn2NiGa}c) is continuously filled. The Fermi energy is consequently shifted to higher band energies and thus $\varepsilon_F$ sweeps over a range of the majority DOSs, whereas the minority spin channel remains unchanged. It is clearly seen from Figs.~\ref{fig:DosMn3Ga} and \ref{fig:DosMn2FeGa} that the tetragonal transition correlates with the peak structure of the majority DOSs. For $\varepsilon_F$ being centered on a peak of the majority DOS, the tetragonal distortion can be triggered. These local maxima are mainly composed of states of the $Y(4c)$ (Mn,~Fe,~Co,~Ni) and Mn $(4d)$ atoms. These energy levels are of $t_{2g}$ symmetry mainly.
In simple interpretation, the DOS can be understood in a rigid-band-like fashion. The limit of this interpretation is reached with Mn$_2$NiGa, where the $d$-PDOS is rearranged and the Slater--Pauling rule is no longer valid for the cubic phase \cite{WCK+2014}.
Comparing this to the PDOS of the tetragonally distorted systems, it is observed that the resulting PDOSs are widely dispersed and significantly less structured. Mn$_2$NiGa, however, behaves differently. Further filling of the majority states, as intuitively expected, does not occur. In contrast, the gap in the minority channel closes as the states of $e_{g}$ symmetry are pulled towards $\varepsilon_F$. Thus, the tetragonal distortion in Mn$_2$NiGa is formed by another mechanism, which may explain why Mn$_2$NiGa is found to be a magnetic shape memory alloy, whereas Mn$_3$Ga and Mn$_2$FeGa are found in their respective tetragonal crystal structures, although the total energy differences are comparable to that of Mn$_2$NiGa.\\
We emphasize that these findings differ from other models in which the instability is thought to depend solely on states of Mn$(4b)$ that is found in a tetrahedral environment. The DOS at $\varepsilon_F$ in mainly composed of states of the $4c$ and $4d$ positions with minor contributions from Mn$(4b)$. Here the instability removes the strong peaks of Mn, Fe, and Co at the $4c$ position, whereas the states of Mn$(4b)$ are not rearranged. 

\subsubsection{Spin-Polarization}
\begin{figure}[htbp]
        \includegraphics[width=\linewidth]{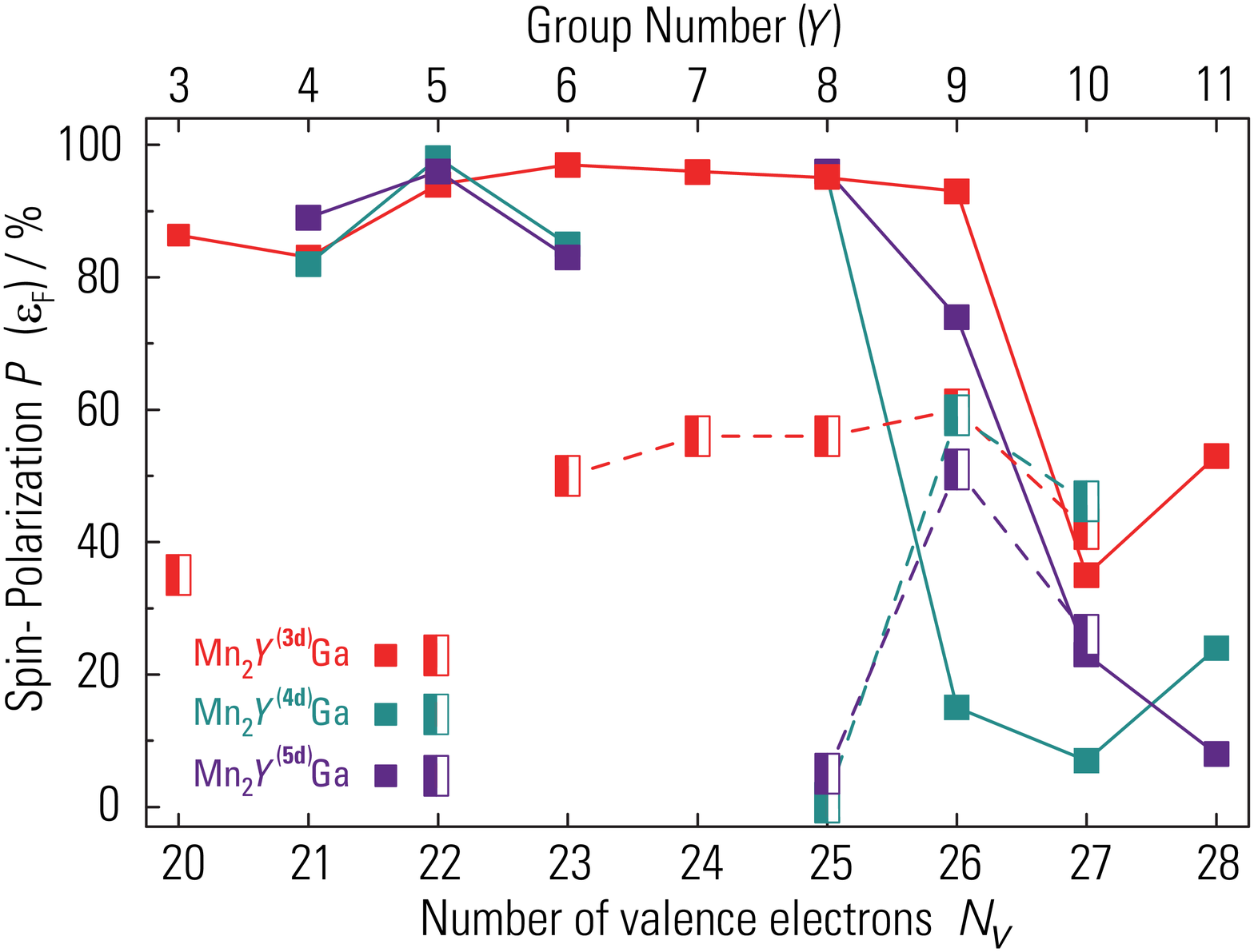}
        \caption{\label{fig:SpinPol}(Color online) The spin-polarization at the Fermi energy for both cubic and tetragonal phases. For a system with $N_{\rm V}< 27$, the spin-polarization is reduced by the tetragonal distortion.}
\end{figure}
Half-metallicity \cite{KWS1983,dGMvE1983} (complete or nearly complete
spin-polarization $P(\varepsilon_{\rm F}) \approx 100\%$) is generally
observed in cubic Co$_2$- and Mn$_2$-based Heusler compounds. The highly
symmetric structure and the peculiar electronic properties due to
covalent bonding lead to the appearance of a gap in the minority density
of states (DOS). The emergence of the tetragonal distortion reduces the spin-polarization of the half-metallic cubic parent phases. The degeneracy of the $t_{2g}$ and $e_g$ states is lifted due to the change in local coordination caused by the distortion, which is seen in Figs.~\ref{fig:DosMn3Ga}--\ref{fig:DosMn2NiGa} and Fig.~\ref{fig:Coord}, and the emergence of a \textsl{pseudo}-gap in one spin-channel is observed instead.
\subsection{Magnetic Ground State}
\label{subsubsec:MagStruct}
\subsubsection{The Slater--Pauling Rule}
\label{subsubsec:SlatPaul}
\begin{figure}[htbp]
\includegraphics[width=\linewidth]{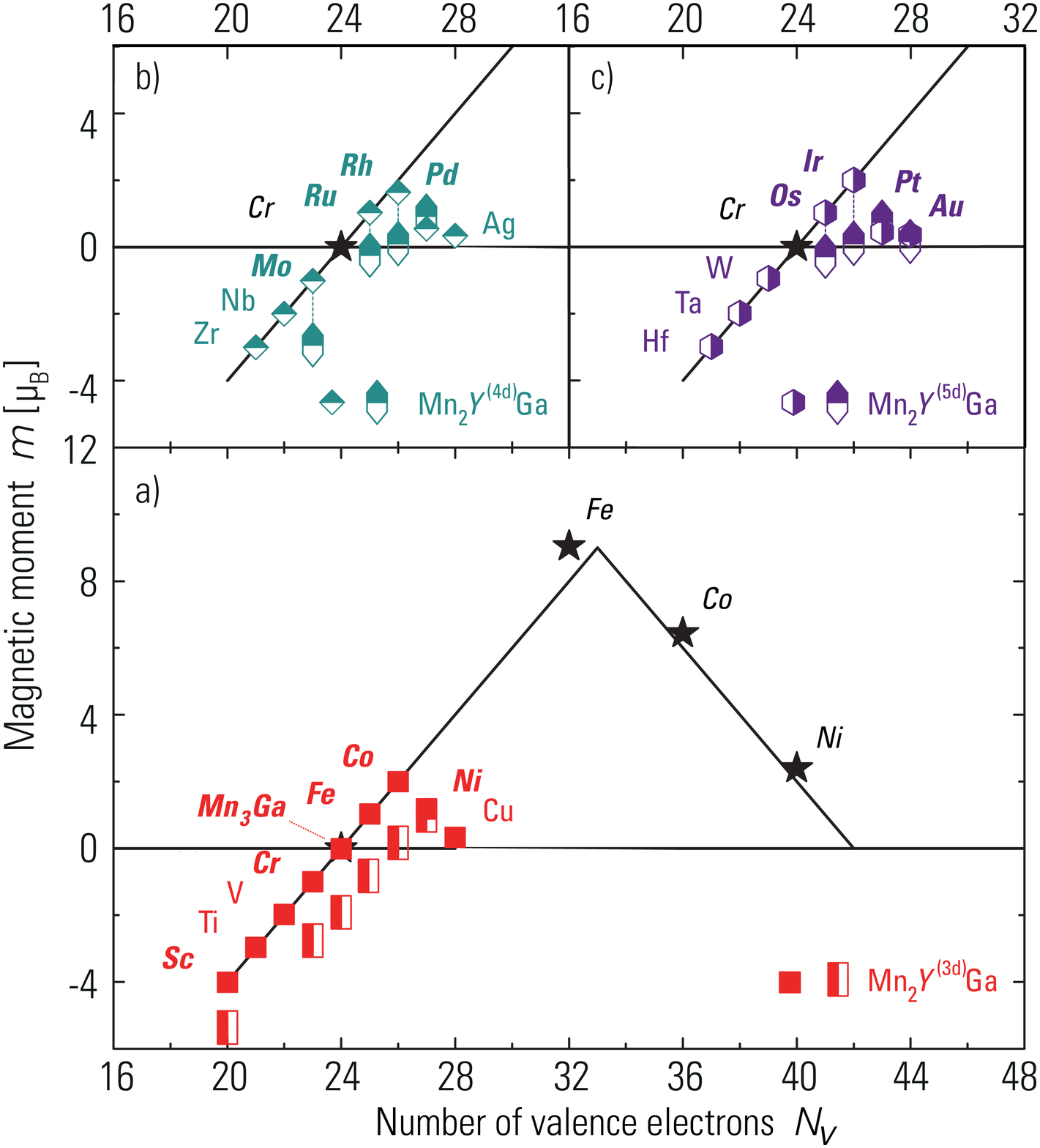}
\caption{\label{fig1-1}(Color online) Slater--Pauling curves of (a) Mn$_2Y^{(3d)}$Ga, (b) Mn$_2Y^{(4d)}$Ga compounds, and (c) Mn$_2Y^{(5d)}$Ga compounds.}
\end{figure}
Referring to a previous publication \cite{WCK+2014} on the cubic parent
compounds of the investigated materials, the results are presented by
means of the Slater--Pauling curves. As can be seen in
Fig.~\ref{fig1-1}, the Slater--Pauling rule experiences strong changes,
so that calling it the Slater--Pauling rule is done for reasons of
convenience. Fig.~\ref{fig1-1} visualizes the fact that the
magnetization of all tetragonal alloys of the Mn$_2Y^{\left(d\right)}$Ga
family experiences a shift to smaller values. In the case of the
Mn$_2Y^{(3d)}$Ga group, this shift is found to be constant throughout
the set of compounds, which results in a linear dependence of the net
moment on the valence electron count, thus giving rise to {\it
  pseudo-Slater--Pauling} behavior, even though half-metallicity and
thus integer net moments are not observed (Fig.~\ref{fig:SpinPol}). A
decrease of the net moment is also found for the Mn$_2Y^{(4d)}$Ga and
Mn$_2Y^{(5d)}$Ga compounds. Unlike the lighter  compounds,  the changes
are not constant over the series and therefore result in nearly
vanishing net moments for Mn$_2$RuGa, Mn$_2$RhGa, Mn$_2$PdGa,
Mn$_2$OsGa, Mn$_2$IrGa, Mn$_2$PtGa, and Mn$_2$AuGa, as is seen in
Fig.~\ref{fig1-1}. Nevertheless, compensation of spin moments may be
achieved for an electron count close to $N_{\rm V}=25.7$, which can be
realized by intermixtures of stoichiometric phases of Mn$_2Y$Ga, such as
${m_{\rm Mn_3Ga}=-1.89~\mu_{\rm B}}$ and a corresponding proportion of
Mn$_2$NiGa with $m_{\rm Mn_2NiGa}=1.00\ \mu_{\rm B}$ or
Mn$_2$CoGa. Thus, fractions of 0.435~$m_{\rm Mn_2NiGa}$ and
0.565~$m_{\rm Mn_2FeGa}$ could ideally lead to complete compensation of
the magnetization. A similar approach was undertaken by \textsl{Nayak et
  al.} that obtained a compensated ferrimagnet by varying the
Mn/Pt-ratio in Mn$_{3-x}$Pt$_x$Ga leading to complete compensation of
magnetization for ${x\approx0.59}$ theoretically~\cite{NNC+2015}. In
Mn$_2$Ru$_x$Ga thin films the compensation of the spin-moment has been achieved through variation of the ruthenium concentration~\cite{KRS+2014}.
\subsubsection{Local Magnetic Moments}
\label{subsubsec:LocMom}
\begin{figure*}[htbp]
\includegraphics[width=\linewidth]{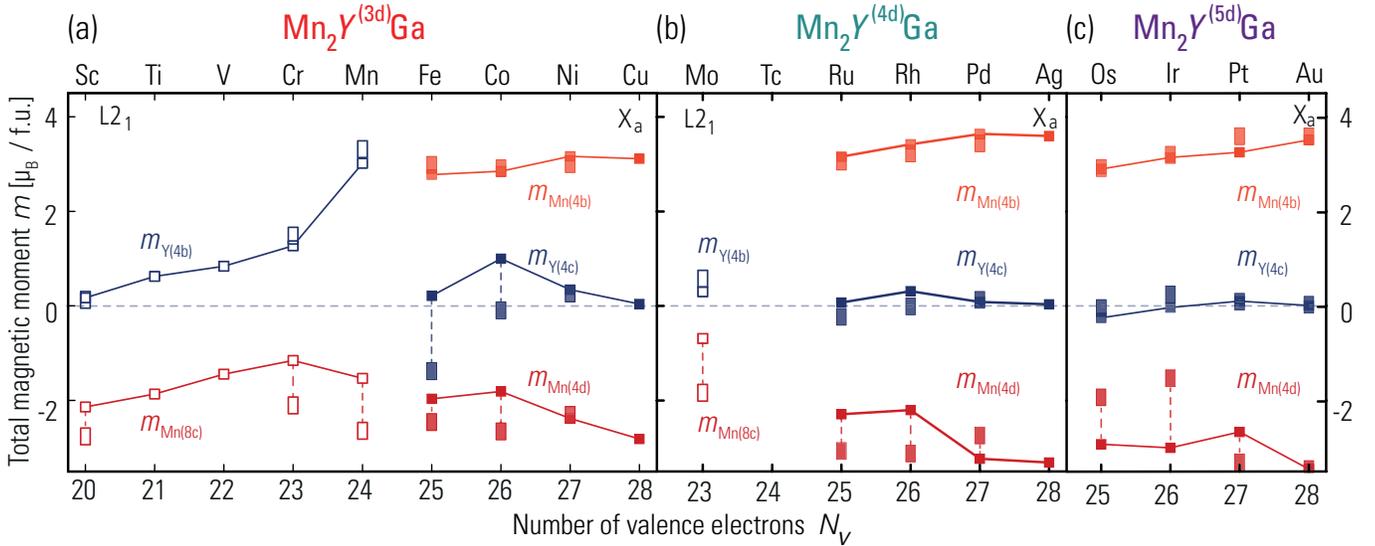} 
\caption{\label{fig:LocMomMn2Y34dGa}(Color online) Atomic magnetic moments  in (a)~Mn$_2Y^{(3d)}$Ga, (b)~Mn$_2Y^{(4d)}$Ga, and (c)~Mn$_2Y^{(5d)}$Ga compounds. Open symbols denote the L2$_1$-type coordination. Filled symbols denote the X$_a$-type coordination. The squares stand for the cubic materials, whereas the tetragonal systems are represented by rectangles.}
\end{figure*}
The change of the total magnetic moment is to be understood in terms of the site moments. Inspection of Fig.~\ref{fig:LocMomMn2Y34dGa} clearly reveals the dependencies. The change in the local moments indicates the specific impact of the distortion on the single sites. 
The local moments responding strongest to elongation/compression of the crystal axes are members of  the magnetic sublattices formed by the Mn$(8c)$, Mn$(4d)$, and $Y(4c)$ sites (the former so-called \textsl{tetrahedral} sites as compared to SG 225). Depending on the characters of the local moments, which are either itinerant or localized  in nature \cite{KWS1983,MSK+2011}, we find major differences in the influence of the tetragonal distortion on these sites. In Fig.~\ref{fig:LocMomMn2Y34dGa}, the localized moment of Mn$(4b)$ is found in the upper part of the plot for a positive value of the magnetic moment of approximately $3~\mu_{\rm B}$.
The effect of elongation along the $c$-axis and compression of the $ab$-plane has a stronger influence on moments of itinerant character found in the lower part of the plot, referring to the $4d$ and $4c$ sites. These are located in the same lattice plane (compare the atoms depicted as red and blue spheres in Fig.~\ref{fig0}). 
Manganese on site $4b$ [Mn$(4b)$] exhibits a large localized moment of
$3~\mu_{\rm B}$, is thus is generally much less affected. Apart from the magnitude of the local moments, Fe$(4c)$ in Mn$_2$FeGa
  exhibits a spin-flip from parallel to anti-parallel alignment of the
  Mn$(4b)$-Fe$(4c)$ interaction upon the tetragonal distortion.
Apart from the changes in magnitude of the local atomic moments, the effective anti-parallel coupling of the nearest neighbor manganese atoms does not suffer from the structural transformation, as will be quantified in terms of the exchange interaction constants in Sec.~\ref{subsec:Exchge}. Since the volume change $C_{\rm t/c}$  is in the order of 1-2\% for the most systems, the nearest-neighbor distance basically does not change, whereas the direction of the nearest-neighbor interaction does (Fig.~\ref{fig:Coord}). In the case of Mn$_3$Ga, $C_{\rm t/c}$ is approximately 9\%, and thus may suppress the shape memory effect in this material.

\subsubsection{Magnetocrystalline Anisotropy}
\label{subsubsec:MagAnis}
\begin{figure}[htbp]
\centering
\includegraphics[width=\linewidth]{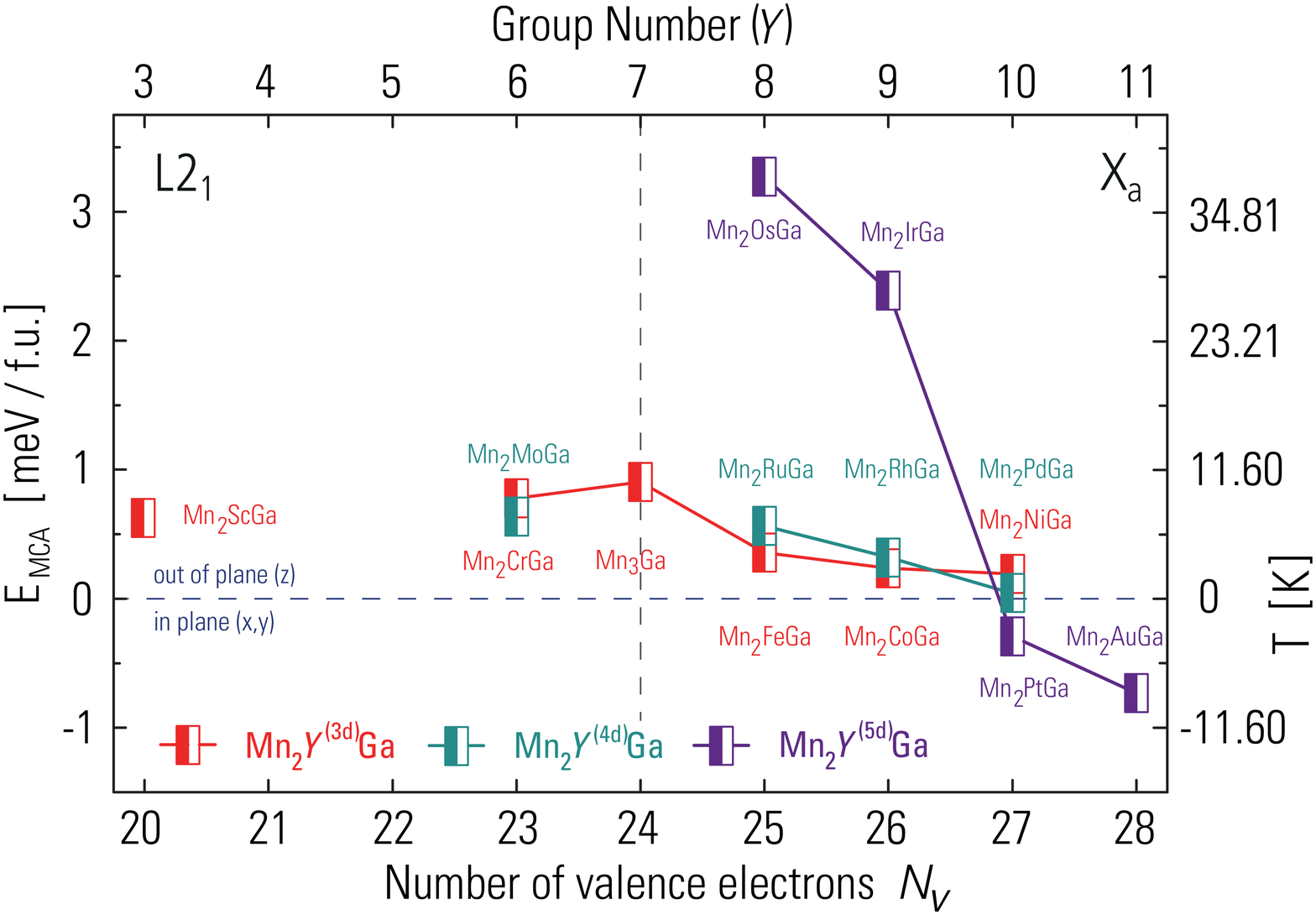} 
\caption{\label{fig:MagAni}(Color online) Calculated MCA energy of Mn$_2Y$Ga.}
\end{figure}

\begin{figure}[htbp]
        \centering
        \includegraphics[width=\linewidth]{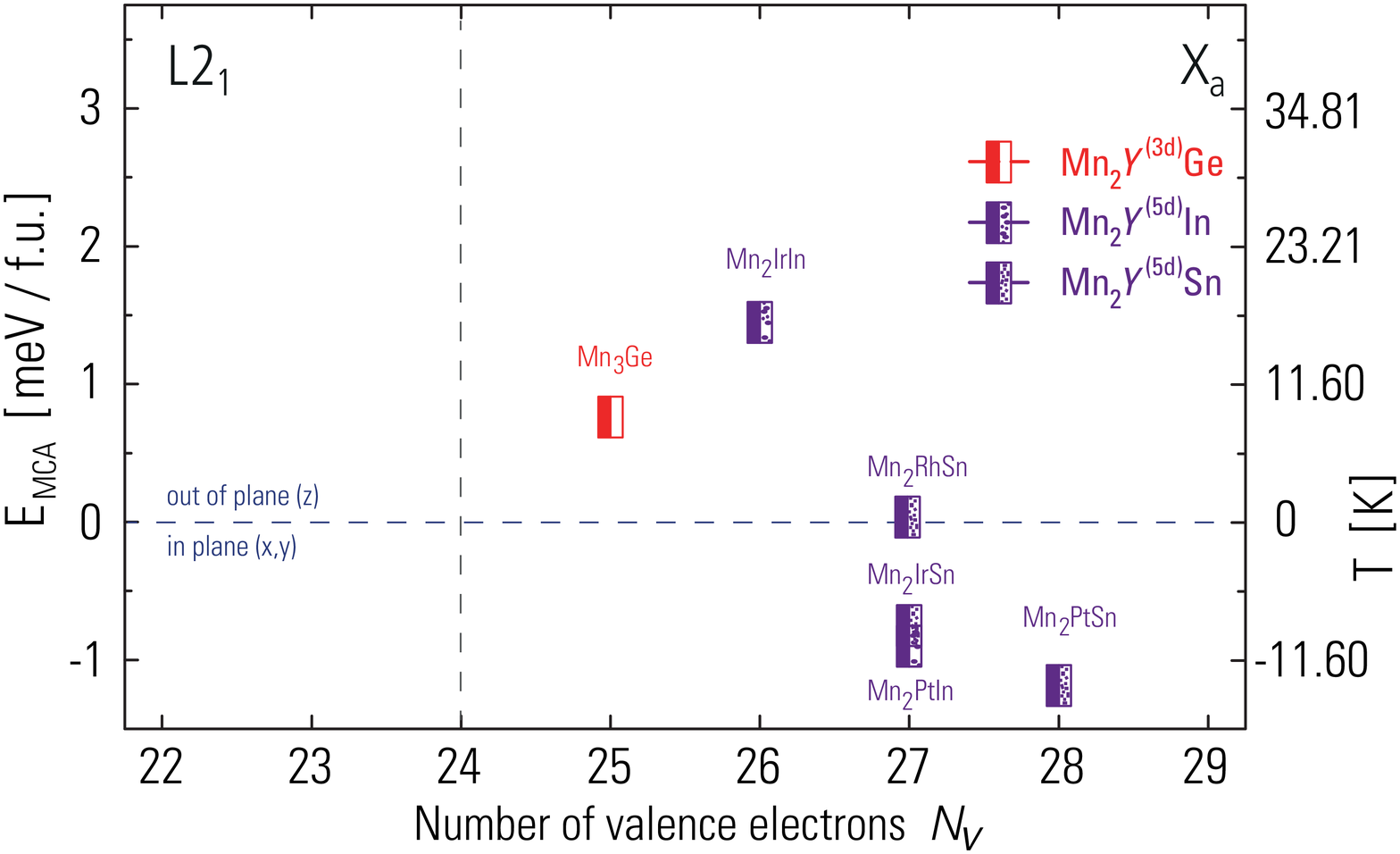} 
        \caption{\label{fig:MagAni2}(Color online) Calculated MCA energy of some chosen Mn$_2YZ$ alloys: Mn$_3$Ge, Mn$_2$IrIn, Mn$_2$IrSn, Mn$_2$PtIn, and Mn$_2$PtSn.}
\end{figure}

Inherent in noncubic crystals is a  directional preference of magnetization that is absent in cubic materials, which is related to the tetragonal modification of the crystal axes. The magnetocrystalline anisotropy (MCA) energy is here  defined as the energy difference between states with magnetization pointing along the \textit{z}-axis and the \textit{x}- or \textit{y}-axis, that is,  $E_{\rm MCA}= E_{\left(100\right)}-E_{\left(001\right)}$, whereas other crystallographic directions are not considered. 
The anisotropy energy is phenomenologically thought to depend on the value of the $c/a$ ratio, which is more or less equal for  most of the compounds investigated in the present study. Therefore, the underlying mechanism is understood as a band-filling effect, which affects the spin orbit coupling (SOC) symmetry. This interpretation can be directly taken from Figs.~\ref{fig:MagAni} and \ref{fig:MagAni2}. Increasing the SOC strength by varying the $Y$ element through the $3d$, $4d$, and $5d$ series increases the MCA energy by a factor of approximately $3$ for $Y=\rm Fe, Ru, Os$. The effect of band filling is deciphered by sweeping the $Y$ elements along a series. Going from left to right in any set of compounds Mn$_2Y^{(xd)}$Ga, the MCA is altered from preferred {\it out-of-plane} to {\it in-plane} orientation. The same situation holds for Mn$_2Y^{(5d)}Z$ compounds, whose preferred orientation is graphed in Fig.~\ref{fig:MagAni2}. Two compounds in this figure deserve special attention, which are Mn$_2$PtIn and Mn$_2$IrSn. Previously they were predicted to possess a non-collinear magnetic order just as Mn$_2$RhSn, which was described in great detail in Ref.~\cite{OCN+2014}. Although we did not consider non-collinear order in the present case, it might be worth while to point out that it is the Mn-atom on site $4d$ that reacts by canting to the incipient spin-reorientation  that is seen clearly in Figs.~\ref{fig:MagAni}. Similar physics is observed in the famous Rare-Earth magnets Nd$_2$Fe$_{14}$B and Er$_2$Fe$_{14}$B. Manganese thus shares properties with Rare Earths~\cite{HLI+2005,YH1986,CGB+1998}.

\subsection{Exchange Coupling and Curie Temperatures}
\label{subsec:Exchge}
\begin{figure}[htbp]
\centering      
        \includegraphics[width=\linewidth]{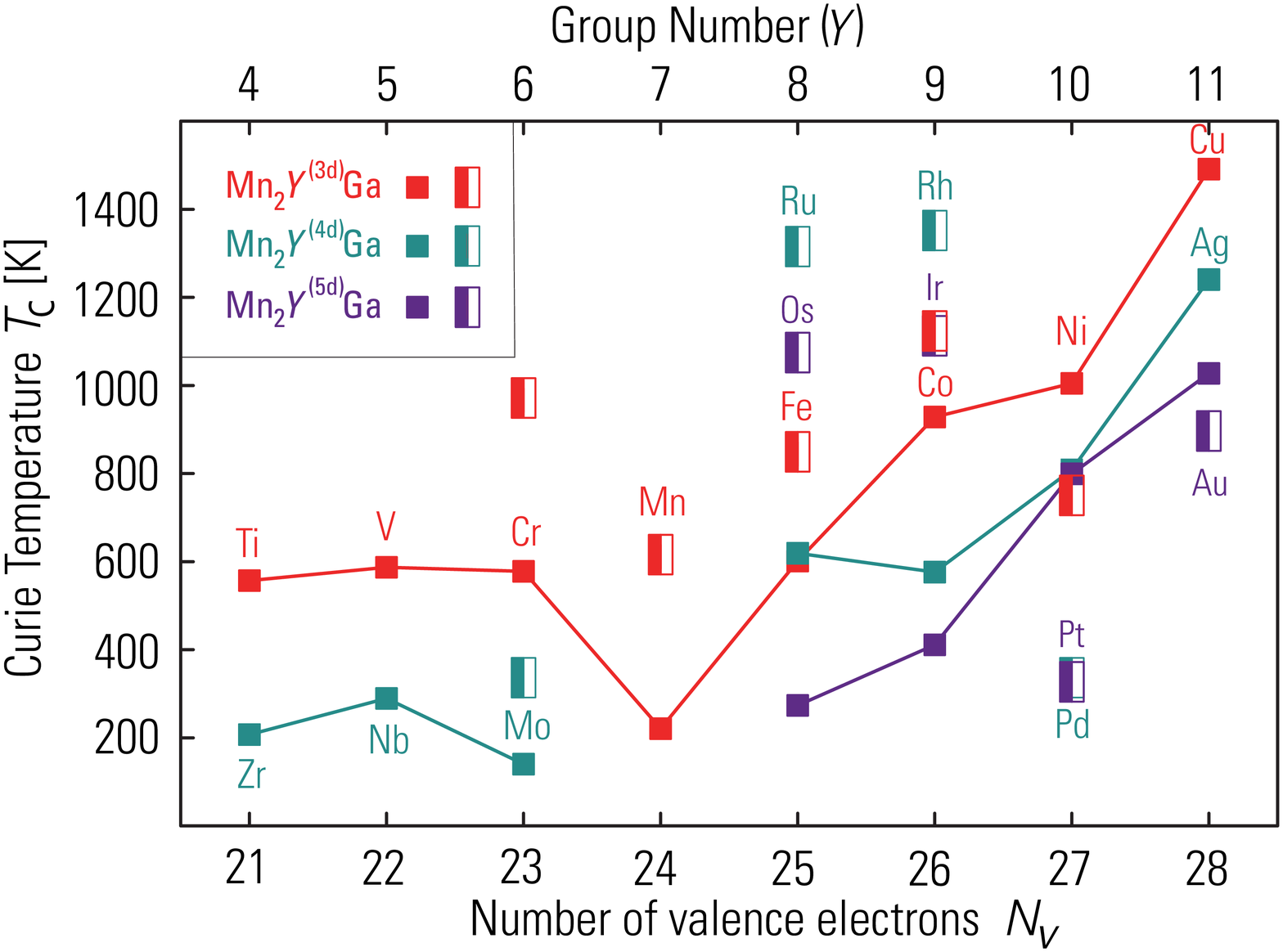} 
        \caption{\label{fig:TC}(Color online) The calculated Curie temperatures of the Heusler compounds containing Ga. The results  shown here  are obtained in the Mean-Field Approximation (MFA) and highlight the consequence of the structural relaxation. Squares correspond to cubic compounds and rectangles to tetragonal compounds.}
\end{figure}

\begin{figure}[htbp]
\centering
        \includegraphics[width=\linewidth]{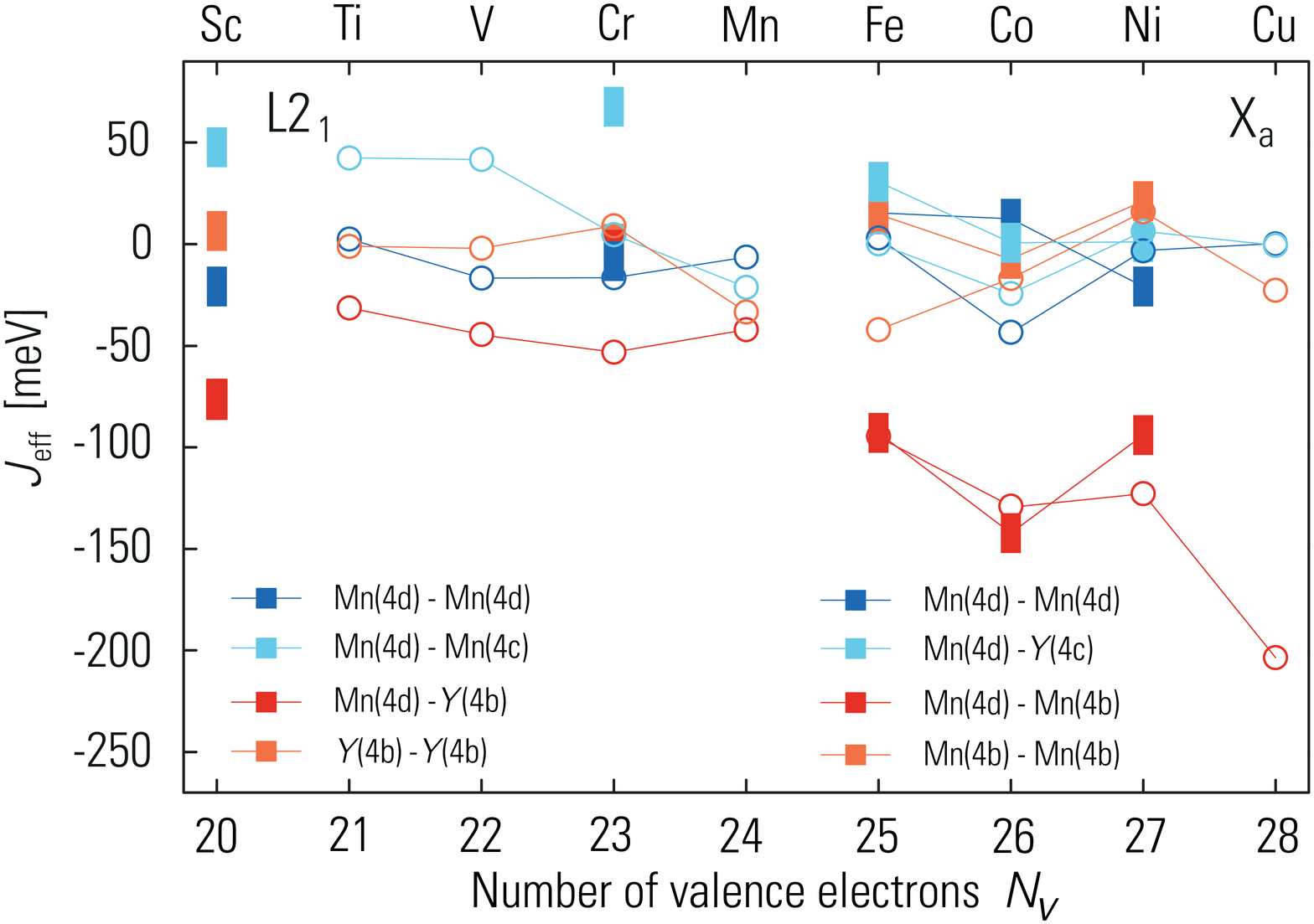} 
        \caption{\label{fig:JIJ}(Color online)  The evaluated effective
          exchange interaction parameters are shown and compared for all
          cubic and tetragonal cases. Thus, the underlying mechanism of
          the increase of $T_{\rm C}$ due to the distortion is visualized. Circles correspond to cubic compounds and rectangles to tetragonal compounds.}
\end{figure}

\begin{figure}[htbp]
        \centering
        \includegraphics[width=.4\linewidth]{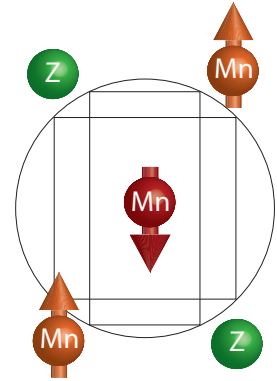} 
        \caption{\label{fig:Coord}(Color online) A two-dimensional projection of the nearest-neighbor coordination of Mn$(4b)$ is shown. The symmetry of the coordination changes while undergoing the {\it martensitic transition}, whereas the nearest-neighbor distance remains unchanged as the volume change is on the order of only 1--9\%.}
\end{figure}

The details of the calculations pertaining to Curie temperatures
($T_{\rm C}$) have been described by us previously~\cite{WCK+2014}. Even
though the in-plane next-nearest-neighbor distance decreases due to the
tetragonal distortion, the next-nearest neighbor Mn$(8c)$-Mn$(8c)$
coupling is still positive for the L2$_1$-derived tetragonal phases;
thus, the  overall magnetic order does not change. A decrease or change
of sign of the coupling constant $J_{{\rm Mn}(4d)-{\rm Mn}(4d)}$ is
typical owing to preferential anti-parallel coupling as caused for short
Mn--Mn distances. Instead, the cubic to tetragonal transition even
results in increased values, thus leading to an increase of $T_{\rm
  C}$. Prima facie, the overall trend of increased $T_{\rm C}$s cannot
be traced back to a common mechanism. In contrast, there is no exchange
interaction that exhibits a similar behavior over the
series. For instance, the findings in the case of
  Mn$_2$FeGa are related to the reduction of magnetic frustration, which
  is due to the competing anti-parallel interaction Mn$(4b)$--Mn$(4d)$, and parallel
  interactions of Fe$(4c)$ with both Mn$(4d)$ and Mn$(4b)$
  neighbors. Upon the tetragonal distortion a spin-flip of the Fe$(4c)$
  local moment is observed, due to the change of the sign of
  Mn$(4b)$--Fe$(4c)$ interaction. Upon the distortion, the strength of
  Mn$(4b)$--Mn$(4d)$ interaction is altered by approximately $\Delta
  J_{\rm Mn(4b)-\rm Mn(4b)}= 56~\rm meV$ in Mn$_2$FeGa. \\
A similar, but   smaller effect is found in Mn$_2$CoGa, indicating the magnetic frustration that had been present in the cubic
phase and the weakening of the exchange interaction $Y(4c)$--Mn$(4b)$
($Y$=~Fe,~Co) due to the tetragonal distortion. This might be one of the
contributions prohibiting the shape memory effect in
Mn$_2$FeGa. The main contribution to $T_{\rm C}$, the Mn$(4d)$--Mn$(4b)$
exchange, does not suffer from the structural transition. 
Similarly, the influence of the distortion on the exchange causes the
smaller contributions to be altered and some frustration to be
diminished. For example, the Mn$(4d)$--Mn$(4d)$ interaction vanishes, 
with preferred anti-parallel alignment in the cubic case,  whereas the major Mn$(4d)$--Mn$(4b)$ interaction remains unchanged.\\
Exceptions to the general observation of increased $T_{\rm C}$ in
tetragonally distorted phases are Mn$_2$NiGa, Mn$_2$PdGa, and Mn$_2$PtGa
systems, in which the $T_{\rm C}$s are reduced upon the tetragonal
distortion. The significant reduction (Fig.~\ref{fig:TC}) is caused by a
weakened Mn$(4d)$--Mn$(4b)$ interaction (Fig.~\ref{fig:JIJ}) that may
indicate an unstable magnetic groundstate. A relation to the Heusler
compound Mn$_2$RhSn~\cite{OCN+2014} can theoretically be established as
these materials possess the same number of valence electrons, $N_{\rm V}$. 
Mn$_2$RhSn has been shown to exhibit a non-collinear magnetic
groundstate due to canting of the different manganese
moments~\cite{OCN+2014}. Non-collinear spin configurations have not been
considered in this work, so that canting of the spins or spin spiral
groundstates in these materials may be present.

\begin{table}[htbp]
        \begin{ruledtabular}
                \begin{tabular}{l|rrrrrD{.}{.}{4}D{.}{.}{2}}
        &       $P_{\rm  c}$    & $P_{\rm  t}$          &$T_{\rm C,c}$  &       \multicolumn{1}{c}{$T_{\rm  C,t}$} &       \multicolumn{1}{c}{$\Delta T_{\rm  C,t-c}$}     & \multicolumn{1}{c}{$E_{\rm  MCA}$} & \multicolumn{1}{c}{$K_u$} \\\hline
        Mn$_2$ScGa      &       87      &       35      &               &       464 &               &       0.616 						&	1.62	\\
        Mn$_2$TiGa      &       83      &               &       557     &               &               &           						&		\\
        Mn$_2$VGa       &       94      &               &       587     &               &               &           						&		\\
        Mn$_2$CrGa      &       97      &       50      &       578     &       970     &       392~&   0.779  						&	2.46	\\
        Mn$_3$Ga        &       96      &       56      &       221     &       610     &       389~&        0.906 						&	2.7	\\
        Mn$_2$FeGa      &       95      &       56      &       601     &       848     &       247~&   0.359  						&	1.16	\\
        Mn$_2$CoGa      &       93      &       60      &       928     &       1124&   196~&   0.236  						&	0.77	\\
        Mn$_2$NiGa      &       35      &       42      &       1005&   750     &       $-$255~&        0.193  						&	0.62	\\
        Mn$_2$CuGa      &       53      &               &       1491&           &               &              						&		\\\hline
        Mn$_2$ZrGa      &       82      &               &       207     &               &               &               						&		\\
        Mn$_2$NbGa      &       98      &               &       289     &               &               &               						&		\\
        Mn$_2$MoGa      &       85      &       65      &       140     &       335     &       196~&   0.636  						&	1.91	\\
        Mn$_2$RuGa      &       95      &       1       &       619     &       1315&   696~&   0.564 						&	1.68	\\
        Mn$_2$RhGa      &       15      &       59      &       576     &       1351&   776~&   0.322  						&	0.95	\\
        Mn$_2$PdGa      &       7       &       46      &       809     &       335     &       $-$473~&   0.040 		 				&	0.11	\\
        Mn$_2$AgGa      &       24      &               &       1240&           &               &              						&		\\\hline
        Mn$_2$HfGa      &       89      &               &               &               &               &               						&		\\
        Mn$_2$TaGa      &       96      &               &               &               &               &               						&		\\
        Mn$_2$WGa       &       83      &               &               &               &               ~&              						&		\\
        Mn$_2$OsGa      &       96      &               &       273     &       1075&   802~&   3.270  						&	9.72	\\
        Mn$_2$IrGa      &       74      &       5       &       411     &       1122&   711~&   2.388  						&	7.02	\\
        Mn$_2$PtGa      &       23      &       51      &       799     &       326     &       $-$472~&        -0.293   						&	-0.84	\\
        Mn$_2$AuGa      &       8       &       26  &   1027&   897     &       $-$130~&        -0.731   						&	-1.94	\\
        
                \end{tabular}
        \end{ruledtabular}
        \caption{\label{tab:TC}The calculated Curie temperatures in Kelvin of tetragonal $T_{\rm C,t}$  and cubic $T_{\rm C,c}$ parent compounds (taken from Ref.~\onlinecite{WCK+2014}). The changes due to the tetragonal transformation are listed as $\Delta T_{\rm  C,t-c}$. $E_{\rm MCA}$ represents the magnetocrystalline anisotropy energies in meV per formula unit, whereas the anisotropy constant, $K_u$, is given in $\rm \frac{MJ}{m^3}$.}
\end{table}

\section{Summary}
\label{sec:Conc}
Using total energy calculations within density functional theory, we
investigated in detail the response to tetragonal distortions for a
large set of cubic Heusler compounds, Mn$_2Y^{(3d,4d,5d)}$Ga, and some
other chosen materials. We were able to single out the systems that
remain cubic from those that favor a tetragonal structure. The details
of the total energy as a function of the distortion were found to be
similar for materials exhibiting the same number of valence
electrons. The magnetizations of the tetragonal alloys were found to be
shifted to smaller values, which we could attribute to changes of the
itinerant local moments. This led to characteristic modifications of the
Slater--Pauling curve. By means of partial densities of states, the
changes to the electronic structures revealed the microscopic origin of
the observed trends. As compared to the cubic parent phases, 
a strengthening of the exchange interaction between neighboring sites
was observed, 
which resulted in an increase of the Curie temperature. Focusing our
attention on the magnetocrystalline anisotropy we observed an
interesting trend that describes a spin reorientation over our series of
compounds, furthermore  very large anisotropies are found for tetragonal
Heusler compounds containing heavy transition metals accompanied by low
magnetic moments, which indicates that these materials may be  promising
candidates for spin-transfer torque applications.

\acknowledgments
Financial support from the Deutsche Forschungsgemeinschaft (DfG) (research unit FOR No.~1464
“ASPIMATT”, project P~1.2-A) and European Research Council Advanced Grant (ERC-AG) No.~291472 ”IDEA Heusler!” is gratefully acknowledged.

\bibliography{bibl}

\end{document}